\documentclass[10pt,journal]{IEEEtran}

\usepackage{epsfig}
\usepackage{times}
\usepackage{float}
\usepackage{afterpage}
\usepackage{amsmath}
\usepackage{amstext}
\usepackage{amssymb,bm}
\usepackage{latexsym}
\usepackage{color}
\usepackage{graphicx}
\usepackage{amsmath}
\usepackage{amsthm}
\usepackage{graphicx}
\usepackage{support-caption}
\usepackage[center]{caption}
\usepackage{pstricks}
\usepackage{graphbox} %loads graphicx package

\usepackage{subcaption}
\usepackage{booktabs}
\usepackage{multicol}
\usepackage{lipsum}% dummy text
\usepackage[T1]{fontenc}
\usepackage{epstopdf}
\usepackage{cite}
\usepackage[hidelinks]{hyperref}
\usepackage{aecompl}
\usepackage{mathrsfs}
\usepackage{multirow}
\usepackage{enumitem} 

\usepackage{tcolorbox}
\usepackage{soul}
\usepackage{mathtools}
\mathtoolsset{showonlyrefs=true}

\newtheorem{thm}{Theorem}%[section]

\newtheorem{lem}{Lemma}

\newtheorem{rem}{Remark}

\theoremstyle{definition}
\newtheorem{defn}{\protect\definitionname}
\providecommand{\definitionname}{Definition}

\newfloat{algorithm}{tbp}{loa}
\providecommand{\algorithmname}{Algorithm}
\floatname{algorithm}{\protect\algorithmname}

\newcommand{\Qsf}{\mathsf{Q}}
\newcommand{\Vsf}{\mathsf{V}}
\newcommand{\Csf}{\mathsf{C}}
\newcommand{\Ncal}{\mathcal{N}}
\newcommand{\Ccal}{\mathcal{C}}
\newcommand{\Rcal}{\mathcal{R}}
\newcommand{\Mcal}{\mathcal{M}}

\allowdisplaybreaks

\usepackage{epstopdf}
\usepackage{pgfplots}
\usepackage{tikz-3dplot}
\tdplotsetmaincoords{60}{115}
\pgfplotsset{compat=newest}

\begin{document}

\title{On Second Order Rate Regions for the \\ Static Scalar Gaussian Broadcast Channel}

\author{
\IEEEauthorblockN{Daniela Tuninetti, Paul Sheldon, Besma Smida and Natasha Devroye}
\thanks{The Authors are with the Electrical and Computer Engineering Department of the University of Illinois Chicago, Chicago, IL, USA. 
E-mails: {danielat, \ psheld2, \ smida, \ devroye @uic.edu}. 
Part of this work was presented at~\cite{9500658_psdtbs_icc2021}. This work was supported in part by NSF Award 1900911.}
}
\maketitle
\begin{abstract}
This paper considers the single antenna, static Gaussian broadcast channel in the finite blocklength regime. Second order achievable and converse rate regions are presented. Both a global reliability requirement and per-user reliability requirements are considered. The two-user case is analyzed in detail, and generalizations to the $K$-user case are also discussed.
The largest second order achievable region presented here requires both superposition and rate splitting in the code construction, as opposed to the (infinite blocklength, first order) capacity region which does not require rate splitting. Indeed, the finite blocklength penalty causes superposition alone to under-perform other coding techniques in some parts of the region. 
In the two-user case with per-user reliability requirements, the capacity achieving superposition coding order (with the codeword of the user with the smallest SNR as cloud center) does not necessarily gives the largest second order region. Instead, the message of the user with the smallest point-to-point second order capacity should be encoded in the cloud center in order to obtain the largest second order region for the proposed scheme.
\end{abstract}


\section{Introduction}
\label{sec:intro}
Wireless communications is deeply integrated into many aspects of everyday life. The delivery on the promise of high bandwidth with reasonable latency has driven much interest into use cases that were previously considered less suitable for wireless communications. These are use cases requiring very low latency coupled with very high reliability. Wireless links are replacing wired links in remote, real-time control and monitoring in manufacturing, and %bringing remote, real time control and monitoring 
in applications where wired links are impossible, such as unmanned aerial vehicles (UAV) and autonomous vehicles. 
For example, a key component of 5G New Radio, Ultra-Reliable and Low Latency Communications (URLLC) is the 5G service category with sub millisecond end-to-end delays and over $99.999\%$ reliability \cite{3gpp.38.913} designed to meet these new requirements. Characterizing the performance of various code constructions operating under URLLC conditions has been a subject of interest~\cite{8594709,7945856}. These works focus on an orthogonal URLLC operation, where communication is modeled as point-to-point links and makes uses of point-to-point results for channels at finite blocklength. However, orthogonalization is known to lead to achievable rates below the capacity of many multi-user channels even in the infinite blocklength case. Thus, understanding the fundamental behavior of multi-user networks at finite blocklengths from an information theoretic standpoint is critical to 
%understand the relative performance of URLLC schemes and
benchmark various neXt URLLC generation (xURLLC) schemes. 
%against what is ultimately possible.

In this paper, we derive approximations to the finite blocklength rate region for the single antenna, static, Gaussian broadcast channel in the spirit of the so-called {\it normal approximation}~\cite{polyanskiy:TIT2010}, which is a refined analysis of how the mutual information density concentrates to its mean as the blocklength increases while the error rate is kept fixed as the blocklength varies. The normal approximation quantifies how many bits can be sent through the channel within a finite number of channel uses while maintaining a given reliability. 
Our proposed scheme uses superposition coding,
%{\blue\cite[Sec. 5.2]{ElGamalKim:book} and \blue\cite[Sec. 15.1.3]{cover}}, 
which achieves the (infinite blocklength, first order) capacity of the considered channel model~\cite[Sec. 5.2]{ElGamalKim:book}.
%which is a well known coding technique for multi user networks in which sub-codebooks are combined to create the complete codebook. 
When decoding for the two-user case, the user with the smallest SNR (referred to as the `weak user') recovers its message while treating the other message as noise. The weak user's message is commonly referred to as the `cloud-center.' The user with the largest SNR (referred to as the `strong user') recovers both messages, and its message is referred to as the `satellite.'
While rate splitting is not needed to achieve the capacity region, it allows one to express the achievable region in a form that can be more easily matched to a converse bound~\cite[Sec. 5.6.1]{ElGamalKim:book}.
{\it Our proposed scheme uses both rate splitting and superposition.}

When considering finite blocklength operation of multi-user networks, care must be taken to how reliability is defined and measured. For the broadcast network, consisting of a single transmitter and multiple receivers, it can take two forms. It may be a {\it global} requirement of reliability, i.e., the joint probability of any user failing to decode its intended message, not exceeding a given value~\cite[Sec 5.1]{ElGamalKim:book}.
%an expected reliability when averaged over all possible codewords. 
Alternatively, it may be a {\it per-user} requirement, where the probability of each user decoding their intended message(s) in error must not exceed a  threshold specified for that user, which may differ across users. 
%
%The global requirement is a suitable model for broadcast networks in which receivers' reliability requirements are undifferentiated.    However, 
In xURLLC, some use cases will have varying reliability requirements. Virtual/Augmented Reality applications will likely have relaxed reliability requirements compared to remote surgical applications.
A transmitter that simultaneously sends entertainment information to one user while transmitting critical public safety information to another is an example. This network should not be constrained by a global error probability, as enforcing the most stringent reliability requirement may significantly reduce the overall performance. %greater error rate in some applications would be acceptable.
{\it This motivates us to consider both definitions of reliability in this work.}

Since the beginnings of information theory as a discipline, much effort has been spent in working to bridge between the elegant convergence of the optimal coding rate to capacity and  results that give more practical insight. In short, what can be said about %the penalty in rate that can be expected for a 
 practical networks that operate at finite blocklength? 
The importance of these non-asymptotic fundamental limits to real networks was recognized very early and the first results were produced almost immediately by Shannon and Feinstein \cite{1057459,SHANNON19576}, and 
%updated in the next decade 
then by Gallager \cite{1053730}. In the ensuing years much progress was made in `large-deviation' analysis, a study of the decay of the probability of increasingly unlikely events. This provided precise values for the rate of decay in the probability of error for fixed rates below capacity as channel uses increased -- the so-called `error exponent regime.'
%It was less understood how the probability of error behaved in response to changes in rate for a fixed number of channel uses.  A precise calculation of error required costly computations of tail bounds, and the accuracy of approximations was either subject to large asymptotic terms or only conjectured.     
%
Hayashi~\cite{Hayashi} and Polyanskiy {\it et al.}~\cite{polyanskiy:TIT2010} improved the state of the art and derived tight non-asymptotic results for a variety of point-to-point channels assuming that the error probability remains fixed while the blocklength increases and the rate converges to capacity -- the so-called `second order regime.' {\it This work adopts the  second order rate region perspective.}

%\subsection{Previous Work on Second Order Regions}
The preceding discussion concerned point-to-point communication problems. The practical usefulness of these results has driven significant interest in applying similar techniques to  multi-user channels.  Much work has focused on the Multiple Access Channels (MAC), such as ~\cite{Huang+Moulin,Tan+Kosut,Laneman-MAC}, which considered both the discrete memoryless and the AWGN models. Interestingly, for the Gaussian MAC the second order region is not tightly characterized yet. 
Other variations on the MAC at finite blocklength have been considered -- such as, fading and random access~\cite{9535162}, the number of users scales with the blocklength~\cite{7529226,MAC_PUPE},  feedback~\cite{8259002}, cooperation~\cite{9140003}, etc. -- but those are not directly relevant to this work.
Directly relevant to our work is~\cite{MACDMS}, which considered the Gaussian MAC {\it with degraded message sets}, that is, one of the two transmitters knows both messages at the time of encoding; in this case the second order region is known. In our conference paper~\cite{9500658_psdtbs_icc2021}, we made use of several techniques developed in~\cite{MACDMS,Laneman-MAC}, such as the multivariate Berry-Essen Theorem and methods for bounding the probability of error for threshold decoding, which we extend here to the case of any number of users and also to the case of per-user reliabilities.

The Broadcast channel (BC) at finite blocklength has been studied for example in~\cite{Tan+Kosut}, where an achievable region for the two-user, discrete memoryless, {\it asymmetric} (where one receiver has to decode both messages) BC was presented; this finds applications in superposition coding methods where one receiver decodes the unintended messages while doing interference stripping decoding.
In~\cite{https://doi.org/10.48550/arxiv.2202.02110},
the two-user AWGN BC with heterogeneous blocklengths was considered; our work with global error is the special case where the two blocklengths are the same, yet our construction produces a larger region in this case.
In~\cite{NOMABC1,NOMABC2} the AWGN BC channel with superposition coding was analyzed based on point-to-point results; it is unclear which code construction would achieve the dispersion utilized in the analysis, possibly that in~\cite{7605463}.

%The techniques borrowed from the MAC literature when combined with insights about the potential for rate splitting and code construction ordering result in larger achievable rate regions than previous AWGN BC results when specialized to our homogeneous blocklength setting. 

Many second order results, including our own, rely on power-shell codebook construction.  A power shell for a codebook of length $n$ is the $(n-1)$ sphere centered at zero whose radius is $\sqrt{nP}$, where $P$ is the average input power constraint. A power shell construction is a random coding argument where codewords are chosen uniformly at random from that $(n-1)$-sphere.
%It is not surprising that power shell construction is seen often in finite blocklength results. 
Power shell construction aligns with Shannon's observation about the optimal decay of the probability of error near capacity of the point-to-point Gaussian channel, which is achieved by codewords on the power-shell~\cite{6767457}.

\subsection{Contributions}
In this paper we aim to characterize the second order rate region of the $K$-user single antenna, static, Gaussian BC, under global and per-user reliability constraints, in the case where the users have the same blocklength. Our main contributions are as follows.
% ACHIEVABILITY
{\bf (1) Achievablity.} By utilizing modified techniques from~\cite{MACDMS}, we show that superposition coding with rate splitting provides the largest second order achievable rate region for this BC network in the case of two users. Through the addition of rate splitting, our achievable region for the two-user case is a super-set of the region presented in~\cite{https://doi.org/10.48550/arxiv.2202.02110} evaluated for equal blocklength for the users. An extension to any number of users, albeit without rate splitting, is also given.
% CONVERSE
{\bf (2) Converse.}  We generalize the converse argument provided in~\cite{https://doi.org/10.48550/arxiv.2202.02110} to the $K$-user case, as well as to the per-user reliability constraints, which to the best of our knowledge has never been reported before.  
% ELSE
{\bf (3) Unexpected behavior under per-user error.} Finally, for the case of per-user reliability and two users, we show that the capacity achieving ordering of superposition coding, where the message for the user with the lowest SNR is encoded in the cloud center, and the message for the user with larger SNR is superimposed as a satellite, does not always achieve the largest second order region. The optimal ordering is instead determined by the second order point-to-point capacities between the transmitter and each of the users. For strictly more than two users, the best superposition coding order with per-user reliabilities changes for different points on the boundary of the second order region.

\subsection{Notation}
%In this paper we use the following notation conventions.
%\begin{itemize}
%  \item %Calligraphic symbols denote sets, bold symbols vectors, and 
%  Sans-serif symbols denote system parameters.
%  \item $|\cdot|$ denotes either the cardinality of a set or the length of a vector.
%  \item $\mathrm{det}(M)$ is the determinant of the matrix $M$.
%  \item $1_{\{\Ec\}}$ is the indicator function of the event $\Ec$.
%  \item $M[\Qc, \Sc]$ is the sub-matrix of $M$ obtained by selecting the rows indexed by $\Qc$  and the columns indexed by $\Sc$.
%  \item For an integer $b$, we let $[b] := \{1, \ldots, b\}$.
%\end{itemize}

For reals $a \leq b$, we let $[a, b] := \{x : a \leq x \leq b\}$.
For integers $a \leq b$, we let $[a: b] := \{a, a+1, \ldots, b\}$ and $[b] := [1: b]$. %, \forall a,b \in \mathbb{Z}^+
$\delta(\cdot)$ is the unit impulse function.
%$O(\cdot)$ is the ``Big O'' notation, which describes the limiting behavior of a function as its argument tends to infinity. 
We write $f(x)=O(g(x))$ if a positive $M$ and an $x_0$ can be found such that  $\vert f(x) \vert \leq Mg(x)$ for all $x\geq x_0$; we also use $O_n$ as a shorthand notation for $O(n)$.
%For a distribution $Q$, $Q^n$ denotes its $n$-fold product.
%For most of the paper we use
%upper-case calligraphic symbols to denote sets,
%upper-case normal symbols for scalar random variables,
%lower-case normal symbols for realizations of random variables, and
%bold-faced symbols for vector or matrices. 
We refer to real-valued vectors of length $n$, either as $x^n$ or $\bm{x}$ (bold font).
$\bm{1}$ and $\bm{0}$ denote the all-one and all-zero vector or matrix, respectively; when needed, their dimension is indicated in the subscript.
For vectors $\bm{a}$ and $\bm{b}$ in $\mathbb{R}^n$, the inner product is denoted as %, \text{ is the inner product in the Euclidean space},
%\begin{align}
$
\langle \bm{a} , \bm{b} \rangle  = \sum_{i\in[n]} a_i b_i,
$
%\label{eq:def n-innerproduct}
%\end{align}
which induces the norm %\text{ is the $\ell_2$ norm}
%\begin{align}
$
\Vert \bm{a} \Vert = \sqrt{\langle \bm{a} , \bm{a} \rangle}.
$
%\label{eq:def n-2norm}
%\end{align}
The $(n-1)$-sphere of radius $r > 0$ is the set 
\begin{align}
\mathcal{S}_{n-1}(r) = \{ \bm{a} \in \mathbb{R}^n: \Vert \bm{a} \Vert = \sqrt{r} \},
\label{eq:def n-shpere}
\end{align}
whose surface area is denoted as  %{\red !!note use of subscript $n$ rather than $n-1$!!}
\begin{align}
S_n(r) = \frac{2 \pi^{n/2}}{\Gamma(n/2)}r^{n-1}.
\label{eq:def n-shpere surface}
\end{align}
Note that the set in~\eqref{eq:def n-shpere} is denoted by the calligraphic font and has subscript $n-1$, while the real non-negative number in~\eqref{eq:def n-shpere surface} is denoted by the normal font and has subscript $n$ as in~\cite{MACDMS}. %, as the $n-1$-sphere is the surface/boundary of an $n$ dimensional ball.

%The scalar Gaussian pdf function with argument $x$, mean $\mu$ and variance $\sigma^2$, is denoted as
%\begin{align}
%\Ncal\left(x; \mu, \sigma^2 \right) = \frac{1}{\sqrt{2\pi\sigma^2}}e^{-\frac{(x-\mu)^2}{2\sigma^2}}.
%\label{eq:def gaussianpfd}
%\end{align}
$\bm{Z} \sim \Ncal\left( \bm{\mu}, \bm{V} \right)$ denotes that $\bm{Z}$ is a jointly Gaussian vector with mean $\bm{\mu}$ and covariance matrix $\bm{V}$, with cumulative distribution function (cdf) 
%of $\bm{Z} \sim \Ncal\left( \bm{0}, \bm{V} \right)$ is denoted as
\begin{align}
\Psi(\bm{x};\bm{\mu}, \bm{V}) 
  = \Pr[\bm{Z} \leq \bm{x}],
\label{eq:def gaussianCDF}
\end{align}
where the inequality ``$\bm{Z} \leq \bm{x}$'' in~\eqref{eq:def gaussianCDF} is intended component-wise, and with probability distribution function (pdf) 
%argument $\bm{x}$ is denoted by
\begin{align}
\Ncal\left(\bm{x}; \bm{\mu}, \bm{V} \right) 
= \frac{\partial \Psi(\bm{x};\bm{\mu}, \bm{V}) }{\partial \bm{x}}
= \frac{{\rm e}^{-\frac{1}{2}(\bm{x}-\bm{\mu})^T\bm{V}^{-1}(\bf{x}-\bm{\mu})}}{\sqrt{\det[2\pi\bm{V}]}}.
\label{eq:def jointlygaussianpfd}
\end{align}
Following~\cite[eq(33)]{MACDMS}, for $\varepsilon\in[0,1]$ and covariance matrix $\bm{V}$, we define  the set
\begin{align}
\Qsf_\text{\rm inv}(\varepsilon;\bm{V}) 
  = \{ \bm{a} : \Psi(-\bm{a};\bm{0},\bm{V}) \geq 1-\varepsilon \}.
\label{eq:def Qinv}
\end{align}

The {\it capacity}, in nats per channel use, of the point-to-point Gaussian channel with SNR $x$ is %(infinite block-length and vanishing error)
\begin{align} 
\Csf(x) 
  = %\frac{1}{2}
  1/2 \, \ln(1+x), \  0\leq x.
\label{eq:def Cx}
\end{align} 
Second order results for multi-user Gaussian channels are often expressed as a function of the {\it cross-dispersion} function %, or variance of the information density
\begin{align}
\Vsf(x,y) 
  = \frac{x(2+y)}{2(1+x)(1+y)}, \  0\leq x \leq y.
\label{eq:def Vxy}
\end{align}
The point-to-point Gaussian {\it dispersion} function is 
\begin{align}
\Vsf(x)
  = \Vsf(x,x) = \frac{x(2+x)}{2(1+x)^2}, \  0\leq x.
\label{eq:def Vx}
\end{align}
%The first two terms in~\eqref{eq:normal approximation M*} are termed the 
The {\it normal approximation} of the second order capacity of the point-to-point Gaussian channel with SNR $x$, for $n$ channel uses and reliability $\varepsilon$, is denoted as
\begin{align}
\kappa(n,x,\varepsilon) 
= \Csf(x) - \sqrt{\frac{\Vsf(x)}{n}} \Qsf^{-1}(\varepsilon), \  
0\leq x, \varepsilon\in[0,1],
\label{eq:def normal approximation p2p}
\end{align} 
%We note that the expression in~\eqref{eq:def normal approximation p2p} is a 
which is an accurate proxy for achievable rates for values of the parameters for which $\kappa(n,\gamma,\epsilon)$ is at least comparable with $\ln(n)/n$~\cite{polyanskiy:TIT2010}.
In~\eqref{eq:def normal approximation p2p}, $\Qsf^{-1}(.)$ denotes the inverse of the function %tail distribution of the standard Gaussian random variable, i.e., of 
\begin{align}
\Qsf(x) = \int_x^{+\infty} \frac{1}{\sqrt{2\pi}}  {\rm e}^{-t^2/2} \, {\rm d}t, \ x \in\mathbb{R}.
\label{eq:def Q}
\end{align}
%and relates to the set $\Qsf_\text{\rm inv}(\epsilon;1)$ as follows
For the scalar case, the set defined in~\eqref{eq:def Qinv} is
%can be expressed as a function of $\Qsf^{-1}(.)$ as
\begin{align}
\Qsf_\text{\rm inv}(\varepsilon;\sigma^2) 
%  &= \{ a : \Pr[Z \leq -a] \geq 1-\epsilon \}
%\\&= \{ a : \Pr[Z \geq a] \geq 1-\epsilon \}
%\\&= \{ a : \Qsf(a) \geq 1-\epsilon \}
%\\&= \{ a : a \leq \Qsf^{-1}(1-\epsilon) \}
%\\&
  = \{ a \in \mathbb{R} : a \leq -\sqrt{\sigma^2} \, \Qsf^{-1}(\varepsilon) \}, \varepsilon\in[0,1].
\label{eq:Qinv relation}
\end{align}
The set in~\eqref{eq:Qinv relation} only contains negative values for $\varepsilon\in[0,1/2)$.

\section{Problem Formulation}
\label{sec:model}

We consider the memoryless $K$-user real-valued static Additive White Gaussian Noise (AWGN) Broadcast Channel (BC), where the channel between the base-station sending signal $X$ and the multiple receivers is modeled as $Y_i = X + Z_i$ for user $i\in[K]$. Here $Z_i$ is the %\sim\Ncal(0,\sigma_i^2)proper-complex 
Gaussian noise at receiver $i$, assumed to be independent of all other noises and of the input, and have zero mean and variance $\sigma_i^2$.
%, and $h_i$ is the channel state at receiver $i$. 
The input $X$ is subject to the power constraint $\mathbb{E}[X^2] \leq P$. 
Given these normalizations, the SNR at receiver~$i$ is $\gamma_i := P/\sigma_i^2, \ i\in[K]$. %|h_i|^2 

We are interested in the 
%case where the base-station must convey information to the users within $n$ channel uses, where $n$ represents a hard deadline for the messages to be received, otherwise they become obsolete and thus useless. We are thus in the realm the
so-called {\it second order regime}, where the block-length $n$ is assumed to be large, but not infinite, and the average probability of error is bounded by $\varepsilon$, which may be small but not vanishing in $n$. For most memoryless point-to-point channels, it has been shown~\cite{polyanskiy:TIT2010, Hayashi} that  $M^*(n,\varepsilon)$, defined as the maximum number of messages that can be sent within $n$ channel uses and with an average probability of error not exceeding $\epsilon$, behaves as 
\begin{align} 
%\frac{1}{n}
1/n \ \ln M^*(n,\varepsilon) = \kappa(n,\gamma,\varepsilon) + O_{\ln(n)/n},
\label{eq:normal approximation M*}
\end{align} 
where the normal approximation function $\kappa(\cdot)$ was defined in~\eqref{eq:def normal approximation p2p}, and where the term $\sqrt{{\Vsf(\gamma)}/{n}} \Qsf^{-1}(\epsilon)$ concisely captures the rate penalty incurred by forcing decoding after $n$ channel uses and allowing a probability of error no larger than $\epsilon\in(0,1)$ on a point-to-point Gaussian channel with SNR $\gamma$.
In this paper we aim to develop expressions akin to~\eqref{eq:normal approximation M*} for the two-user AWGN BC.  We will also provide extensions to any number of users. %in Sections~\ref{sec:outK} and~\ref{sec:inK}.
We start with the formal definition of the second order region for the two-user case, which can be straightforwardly extended to any number of users.

\begin{defn}[Code with Global Error]\label{def:codeglobal}
Given integer sets $(\Mcal_0,\Mcal_1,\Mcal_2)$, integer $n$, and non-negative reals $(P,\epsilon)$, an $(n,|\Mcal_0|,|\Mcal_1|,|\Mcal_2|,P,\epsilon)$ code for the two-user AWGN BC has:
(i) three independent and uniformly distributed messages on $\Mcal_0 \times \Mcal_1 \times \Mcal_2$;
(ii) one encoder function $\mathsf{enc}: \Mcal_0 \times \Mcal_1 \times \Mcal_2 \to \mathbb{R}^n$ with %maximal per-codeword 
power constraint
\begin{align}
\Vert \mathsf{enc}(m_0,m_1,m_2) \Vert^2 \leq n P, 
\label{eq:per codeword power constraint}
\end{align}
for all $(m_0,m_1,m_2) \in \Mcal_0 \times \Mcal_1 \times \Mcal_2$; and
(iii) two decoder functions $\mathsf{dec}_k: \mathbb{R}^n \to \Mcal_0 \times \Mcal_k, k\in[2]$, with average global probability of error satisfying
\begin{align}
\Pr\big[ \cup_{k\in[2]} \mathsf{dec}_k(Y_k^n) \not= (W_0,W_k) \big] \leq \varepsilon,
\label{eq:PenGlobal}
\end{align}
where in~\eqref{eq:PenGlobal} it is understood that $(W_0,W_1,W_2)$ was sent.
\hfill$\square$
\end{defn}

We shall use $\varepsilon$ to denote the largest allowed average probability of error, and $\epsilon_n$ for the probability of error of a code of block-length $n$. Again note the difference in font type.

\begin{defn}[Second Order Capacity Region with Global Error]\label{def: Second Order Region}
A non-negative rate tuple $(R_0,R_1,R_2)$ is said to be $(n,\varepsilon)$-achievable if there exists a $(n,M_{0,n},M_{1,n},M_{2,n},P,\epsilon_n)$ code with global error for some $n$ with $\epsilon_n \leq \varepsilon$ and $\frac{\ln(M_{j,n})}{n} \geq R_j $ for $j\in\{0,1,2\}$. Let $\Ccal(n,\varepsilon)$ denote the set of all $(n,\varepsilon)$-achievable rate tuples, referred to as the {\it second order capacity region (with global error)}.
\hfill$\square$
\end{defn}

\begin{defn}[Capacity Region]\label{def:Capacity Region}
The capacity region $\Ccal$ is %defined as %(first order) 
\begin{align}
\Ccal(\varepsilon) &= \cup_{n\geq 1} \Ccal(n,\varepsilon), &&\text{\rm($\varepsilon$-capacity region)},
\label{eq:def C(Eps,n)}
\\ 
\Ccal &= \cap_{\varepsilon>0} \Ccal(\varepsilon),          &&\text{\rm(capacity region)}.
\label{eq:def C(Eps,infty)}
\end{align}
The two-user Gaussian BC enjoys a strong converse~\cite{book:ElGamal-Kim}, that is, the capacity region satisfies (where WLOG $\gamma_1 \geq \gamma_2$)
\begin{subequations}
\begin{align}
\Ccal = \Ccal(\varepsilon) 
&= \bigcup_{\alpha\in[0,1]}
\Big\{ (R_0,R_1,R_2)\in \mathbb{R}^3_+ :
\\ R_0+R_2 &\leq \Csf\left(\frac{(1-\alpha)\gamma_2}{1+\alpha \gamma_2}\right),
%\Csf\left( \gamma_2 \right) - \Csf\left( \alpha \gamma_2 \right), %&&\text{\rm(Receiver 2 is the weak receiver)} %= \Csf\left( \frac{(1-\alpha) \gamma_2}{1+\alpha \gamma_2} \right)
\\  R_1 &\leq \Csf\left( \alpha \gamma_1 \right) \Big\}.                         %&&\text{\rm(Receiver 1 is the strong receiver)}
\end{align}
\label{eq:2BCAWGNCapacity}
\end{subequations}
where $\alpha$ is interpreted as the power split parameter.
\hfill$\square$
\end{defn}

\noindent
{\bf Goal}. We aim to find, or bound, the second order region $\Ccal(n,\varepsilon)$ 
%in Definition~\eqref{def: Second Order Region} 
by characterizing the rate penalty terms to be included in the capacity region in~\eqref{eq:2BCAWGNCapacity} akin to the term $\sqrt{{\Vsf(\gamma)}/{n}} \Qsf^{-1}(\epsilon)$ %in the normal approximation
in~\eqref{eq:def normal approximation p2p} for point-to-point channels.
%, for sufficiently large $n$,

\begin{rem}[On Per-User Error]
\rm
\label{def:codeperuser}
We shall also use, instead of the global probability of error in~\eqref{eq:PenGlobal}, the per-user average error probability criteria
\begin{align}
\Pr[  \mathsf{dec}_k(Y_k^n) \not= (W_0,W_k)
%\ (W_0,W_1,W_2)~\text{sent}
] \leq \varepsilon_k, \quad  k\in[2].
\label{eq:PenPerUser}
\end{align}
The definition of code and second order region with per-user error in~\eqref{eq:PenPerUser} follow similarly to those with global error and is not repeated here for sake of space. 
\hfill$\square$
\end{rem}

\section{Main Result}
\label{sec:mainresult}
The main result of this paper for the two-user case is summarized in Theorem~\ref{thm:main}.
The converse proof can be found in Section~\ref{sec:converse} and the achievability in Section~\ref{sec:achievability}.
Extensions to the $K$-user case can be found in Sections~\ref{sec:outK} and~\ref{sec:inK}.

\begin{thm}[Second Order Regions with Global Error]
\label{thm:main}
Given the model in Section~\ref{sec:model} for global error $\varepsilon$, we have \begin{align}
\Rcal^{\text{\rm(SUP)}}(n,\varepsilon) \subseteq \Ccal(n,\varepsilon) \subseteq \Rcal^{\text{\rm(CS)}}(n,\varepsilon),
\label{eq:main result}
\end{align}
where the regions $\Rcal^{\text{\rm(SUP)}}(n,\varepsilon)$ and $\Rcal^{\text{\rm(CS)}}(n,\varepsilon)$ are as follows.

The region $\Rcal^{\text{\rm(SUP)}}(n,\varepsilon)$ is attained by {\bf superposition coding with rate splitting} and is given by
\begin{subequations}
\begin{align}
\Rcal^{\text{\rm(SUP)}}(n,\varepsilon) 
&=
\hspace*{-1cm}
\bigcup_{(\alpha,\beta,\epsilon_{10},\epsilon_{11},\epsilon_2)\in[0,1]^5} % \text{\rm eq}\eqref{eq:normal approximation sup all}}%_{\alpha\in[0,1]} 
\hspace*{-.3cm}
\Big\{ (R_0,R_1,R_2)\in \mathbb{R}^3_+ :
\\ R_0+R_2+\beta R_1 &\leq \Csf\left(\frac{(1-\alpha)\gamma_2}{1+\alpha \gamma_2}\right)
\\  &\hspace*{-1cm} -\sqrt{\frac{1}{n} \Vsf^\prime(\alpha \gamma_2, \gamma_2)} \Qsf^{-1}\left( \epsilon_2 \right) + O_{\ln(n)/n},
\label{eq:normal approximation sup cloudcenter}
\\ (1-\beta)R_1      &\leq \kappa(n,\alpha\gamma_1, \epsilon_{10}) + O_{\ln(n)/n},
\label{eq:normal approximation sup satellite}
\\ R_0+R_1+R_2       &\leq \kappa(n,      \gamma_1, \epsilon_{11}) + O_{\ln(n)/n}
\label{eq:normal approximation sup sumrate}
\Big\},
\end{align}
\label{eq:normal approximation sup}
where $\alpha$ is the power split and $\beta$ the rate split. 
\end{subequations}
\begin{subequations}
The dispersion in~\eqref{eq:normal approximation sup cloudcenter} is defined as
\begin{align}
  \Vsf^\prime(\alpha \gamma_2, \gamma_2) 
   &:=\Vsf(\alpha\gamma_2) +  \Vsf(\gamma_2) -2 \Vsf(\alpha \gamma_2, \gamma_2)
\\&=
%V_{11} := 
\frac{(1-\alpha)\gamma_2(2\alpha \gamma_2^2+\gamma_2+3\alpha \gamma_2+2)}{2(\gamma_2+1)^2(\alpha \gamma_2+1)^2},
%%%%\leq  \Vsf\left(\frac{(1-\alpha)\gamma_2}{1+\alpha \gamma_2}\right), 
%
%plot 1/2*x*(5)*(x*(5)+2)/(x*(5)+1)^2 for x=0 .. 1
%plot (1-x)(2*x*(5)^2 + (5) + 3*x*(5) + 2)/(x*(5)+1)^2 *(5)/(1+(5))^2/2 for x=0 .. 1
%plot (sqrt(1/2*(5)*((5)+2)/((5)+1)^2) - sqrt(1/2*x*(5)*(x*(5)+2)/(x*(5)+1)^2))^2 for x=0 .. 1
%plot 1/2*((1-x)*(5)/(1+x*(5)))*(((1-x)*(5)/(1+x*(5)))+2)/(((1-x)*(5)/(1+x*(5)))+1)^2) for x=0 .. 1
%
%%%\\&V_{22}  = \frac{( 2 + \alpha \gamma_1) \alpha \gamma_1}{2 (\alpha \gamma_1+1)^2} = \Vsf(\alpha \gamma_1), 
%%%%plot ( 2 + x ) x / (x +1)^2
\end{align}
\label{eq:def V11}
\end{subequations}
with $\Vsf(\cdot,\cdot)$ and $\Vsf(\cdot)$ are defined in~\eqref{eq:def Vxy} and~\eqref{eq:def Vx}, respectively.
The triplet $(\epsilon_{10},\epsilon_{11},\epsilon_2)\in[0,1]^3$ satisfies 
\begin{align}
&(1-\epsilon_1) (1-\epsilon_2) \geq 1-\varepsilon,
\label{eq:sup globalerror}
\end{align}
where $\epsilon_1$ is the error rate at receiver~1 which satisfies
\begin{align}
&\mathsf{F}(\epsilon_{10},\epsilon_{11};r(\alpha\gamma_1, \gamma_1)) \geq 1-\epsilon_1,
\label{eq:sup globalerror at Rx1}
\end{align}
where the probability of correct decoding function $\mathsf{F}(\cdot,\cdot;\cdot)$ is %defined as 
\begin{subequations}
\begin{align}
\mathsf{F}(\epsilon_{10},\epsilon_{11};r) &:=  
\Pr\big[G_2 \leq \Qsf^{-1}(\epsilon_{10}),
    \\&r G_2 + \sqrt{1-r^2} G_3 \leq \Qsf^{-1}(\epsilon_{11})   \big],
\end{align}
\label{eq:sup Fdef} 
\end{subequations}
for $G_2,G_3$ i.i.d. standard Gaussian random variables, and the correlation coefficient $r(\alpha\gamma_1, \gamma_1)$ in~\eqref{eq:sup globalerror} is defined as
\begin{align}
  &r(\alpha\gamma_1, \gamma_1) 
  :=\frac{\Vsf(\alpha \gamma_1, \gamma_1)}{\sqrt{\Vsf(\alpha\gamma_1) \Vsf(\gamma_1)}}
%%%=\frac{V_{23}}{\sqrt{V_{22}V_{33}}} 
= \sqrt{\frac{( 2 + \gamma_1) \alpha}{ ( 2 + \alpha \gamma_1) } }.
\label{eq:def r}
\end{align}
%with $\Vsf(\cdot,\cdot)$ and $\Vsf(\cdot)$ are defined in~\eqref{eq:def Vxy} and~\eqref{eq:def Vx}, respectively.

The region $\Rcal^{\text{\rm(CS)}}(n,\epsilon)$ is the {\bf cut-set}-type region
\begin{subequations}
\begin{align}
\Rcal^{\text{\rm(CS)}}(n,\varepsilon)
&=\Big\{(R_0,R_1,R_2)\in \mathbb{R}^3_+ :
\\ R_0+R_1     &\leq \kappa(n,\gamma_1,\varepsilon) + O_{\ln(n)/n},
\\ R_0+R_2     &\leq \kappa(n,\gamma_2,\varepsilon) + O_{\ln(n)/n},
\\ R_0+R_1+R_2 &\leq \kappa(n,\max(\gamma_1,\gamma_2),2\varepsilon) + O_{\ln(n)/n}
\Big\}.
\end{align}
\label{eq:cutset globalerror}
%evaluated for $$\varepsilon_{\red 1}=\varepsilon_{\red 2}=\varepsilon$.
\end{subequations}
\end{thm}

\begin{rem}[Second Order Regions with Per-User Error]\rm
In Theorem~\ref{thm:main}, the achievable second order region $\Rcal^{\text{\rm(SUP)}}(n,\varepsilon)$ in~\eqref{eq:normal approximation sup} without the constraint in~\eqref{eq:sup globalerror}, which links the error rates at the two receivers (that experience independent noise by assumption), gives an achievable region for the case with per-user error criteria. When we remove the constraint in~\eqref{eq:sup globalerror}, we indicate the achievable region as $\Rcal^{\text{\rm(SUP)}}(n,\epsilon_1,\epsilon_2)$ to stress the two per-user probability of error requirements.

With per-user error, the achievable region akin to the one in Theorem~\ref{thm:main} is $\Rcal^{\text{\rm(SUP1)}}(n,\epsilon_1,\epsilon_2) \cup \Rcal^{\text{\rm(SUP2)}}(n,\epsilon_1,\epsilon_2)$, where $\Rcal^{\text{\rm(SUP2)}}(n,\epsilon_1,\epsilon_2)$ is the region in~\eqref{eq:normal approximation sup} (with the superposition coding order that is capacity achieving under the assumption $\gamma_1 \geq \gamma_2$), and the region $\Rcal^{\text{\rm(SUP1)}}(n,\epsilon_1,\epsilon_2)$ is similar to the region in~\eqref{eq:normal approximation sup} but with the role of the users swapped (that is, with the message of user~1 in the cloud center). While swapping the order of superposition coding does not appear to enlarge the achievable region in Theorem~\ref{thm:main} for global error, it provides improvements when one considers per-user error as we will show in 
Section~\ref{sec:numericalEvaluations}.

The outer bound region $\Rcal^{\text{\rm(CS)}}(n,\varepsilon)$ in Theorem~\ref{thm:main} can also be extended to the case of per-user error. In particular, the single user bounds read $ R_0+R_j \leq \kappa(n,\gamma_j,\varepsilon_j) + O_{\ln(n)/n}$ for $j\in[2],$ and the sum-rate bound becomes $ R_0+R_1+R_2 \leq \kappa(n,\max(\gamma_1,\gamma_2),\varepsilon_1+\varepsilon_2) + O_{\ln(n)/n}$.
\hfill$\square$
\end{rem}

\begin{figure}
     \centering     \includegraphics[width=0.86\columnwidth]{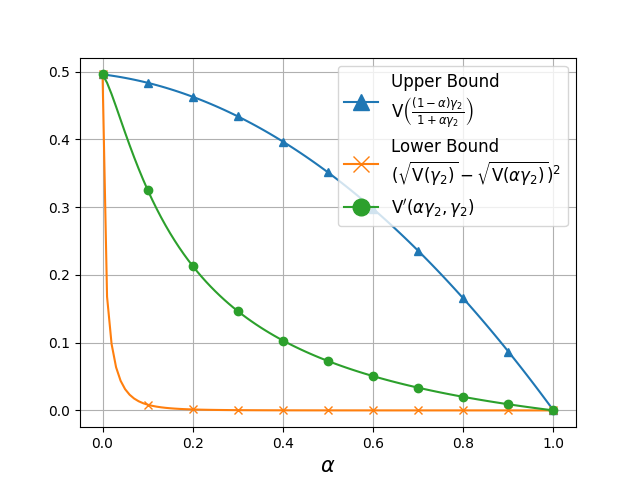}%{dispersionsVSalpha}
     \caption{\small Dispersions vs. $\alpha$ for $\gamma_2=10$. }
     \label{fig:dispersions}
\end{figure}

\begin{rem}[On the Dispersion of Decoding the Message in the Cloud Center]\label{rem:onVprime}
\rm 
Let
\begin{align}
x:=\alpha \gamma_2 \leq  y:=\gamma_2, \ \  z:=\frac{y-x}{1+x}=\frac{(1-\alpha)\gamma_2}{1+\alpha\gamma_2},
\end{align}
%and $\Vsf(\cdot,\cdot)$ as defined in~\eqref{eq:def Vxy}. Note that 
where $z$ represents the SINR in decoding the cloud center by treating the satellite as a noise in Theorem~\ref{thm:main}. The dispersion $\Vsf^\prime(\cdot,\cdot)$ in~\eqref{eq:def V11} can be upper bounded as follows  
\begin{subequations}
\begin{align}
\Vsf^\prime(x,y) 
  &= \Vsf(x)+\Vsf(y)-2\Vsf(x,y) 
\\&= \frac{(y-x)(2xy+3x+y+2)}{2(1+x)^2(1+y)^2} % \underbrace{(2xy+3x+y+2)}_{=2(x+1)^2+(y-x)(2x+1)}
\\&= \frac{z(2+z \frac{2x+1}{x+1})}{2(1+z)^2(1+x)}
%\\&\leq \Vsf(z) \left( 1 - \left(\frac{x}{1+x}\right)^2 \right) 
\leq \Vsf(z),
\label{eq:on V11 upper}
\end{align}
and lower bounded as follows  
\begin{align}
\Vsf^\prime(x,y)
  &\geq \Vsf(x)+\Vsf(y)-2\sqrt{\Vsf(x)\Vsf(y)}
\\&= (\sqrt{\Vsf(y)} - \sqrt{\Vsf(x)})^2.
\label{eq:on V11 lower}
\end{align}
\label{eq:on V11}
\end{subequations}
Recall that $\Vsf^\prime(\cdot,\cdot)$ in~\eqref{eq:def V11} is the dispersion for the rate of messages carried by the cloud center. 
%(see~\eqref{eq:cloud center}). in~\eqref{eq:satellite} 
From the upper bound in~\eqref{eq:on V11 upper}, we see that $\Vsf^\prime(\cdot,\cdot)$ in our scheme is lower than the dispersion of a point-to-point Gaussian channel in which the interference from the satellite codeword is treated as Gaussian noise. 
We do not have at present an intuitive interpretation of the lower bound in~\eqref{eq:on V11 lower}.
%%%======BEG========
%%%The lower bound in~\eqref{eq:on V11 lower} can be interpreted as: the dispersion $\Vsf^\prime(\cdot,\cdot)$ is larger than that of MAC+DMS \nbl{acronym not defined} receiver that decodes the cloud center first and the satellite codeword next, that is, jointly decoding the cloud center and the satellite would lead to a lower dispersion for the \st{could} cloud center than treating the satellite as noise {\red DO U AGREE ON INTERPRETATION?}.{\magenta PS: I am not sure I agree.  The statement seems a bit self contradictory in that the MAC+DMS receiver is decoding the cloud center and then the satellite, but then you say jointly decoding leads to a lower dispersion for the cloud center. I do think the final conclusion makes sense as I would hope that once we have decoded the combined message and the satellite I would think there would be vary little variability compared to a deterministic channel of the cloud center capacity}
%%%======END========
The dispersion $\Vsf^\prime(\alpha \gamma,\gamma)$ vs. $\alpha$ is depicted in Fig.~\ref{fig:dispersions}.
%where we see clearly that the upper bound in~\eqref{eq:on V11 upper} which was used for example in~\cite{NOMABC1,NOMABC2}, can be much larger than what our superposition coding scheme attains.
In~\cite[Theorem 2]{7605463} the Authors considered the performance of nearest-neighbor decoding of independent codewords drawn uniformly at random from two classes of distributions.  We note that $\Vsf^\prime(\cdot,\cdot)$ in~\eqref{eq:def V11} is the special case of~\cite[Eq(23)]{7605463} for codes on the power sphere for the AWGN channel with two users. The same paper also shows that with i.i.d. Gaussian codes, on the AWGN channel, and with nearest-neighbor decoding, the dispersion is~\cite[Eq(27)]{7605463}, which equals $z/(1+z)$ where $z$ is the SINR. The dispersion $z/(1+z)$ is often used to assess NOMA performance by means of (sub-optimal)  point-to-point results.
%was used in~\cite{NOMABC1, NOMABC2} but without any formal introduction of the code construction and decoding strategy adopted.
%%%======BEG========
%%%{\red DO THEY SAY WHAT CODE AND DECODING TEHY USE? NO, RIGH? BUT THEY WRITE $1-(1+P)^{-2}$?}. {\magenta in \cite{NOMABC2} Section III.B claims iid complex Gaussion with a power split between $x_1,x_2$. They claim the paper ``Quasi-Static Multiple-Antenna Fading Channels at Finite Blocklength'' shows $\Vsf$, but that paper is not multiple user. \cite{NOMABC1} makes no justification for their $\Vsf$ and do not discuss how the code is constructed.}
%~\cite{NOMABC1, NOMABC2},
%Yu, H. Chen, Y. Li, Z. Ding, and B. Vucetic, “On the performance of
%non-orthogonal multiple access in short-packet communications,” IEEE
%Communications Letters, vol. 22, no. 3, pp. 590–593, 201
%
%X. Sun, S. Yan, N. Yang, Z. Ding, C. Shen, and Z. Zhong, “Short-
%packet downlink transmission with non-orthogonal multiple access,”
%IEEE Transactions on Wireless Communications, vol. 17, no. 7, pp.
%4550–4564, 2018
%%%======BEG========
\hfill$\square$
\end{rem}

\begin{figure}
     \centering
\includegraphics[width=0.86\columnwidth]{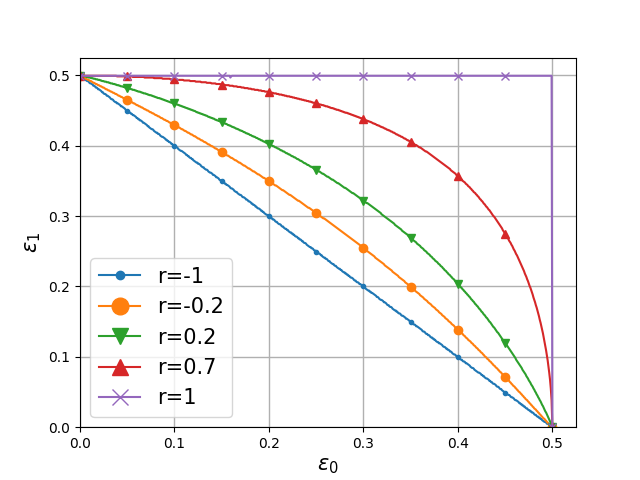}%{epsilonContours}
     \caption{\small Region $\{ (\epsilon_{0},\epsilon_{1})\in[0,1]^2 : \mathsf{F}(\epsilon_{0},\epsilon_{1};r) \geq 1-\varepsilon=0.5 \}$ for various values of $r$.}
     \label{fig:epsregion}
\end{figure}

\begin{rem}[On Reliability Allocation]\label{rem:onF}
\rm 
The probability of correct decoding function in~\eqref{eq:sup Fdef} is monotonic in the correlation coefficient $r\in[-1,1]$.
Some of its values are %can be explicitly computed as
\begin{subequations}
\begin{align}
\mathsf{F}&(\epsilon_{0},\epsilon_{1};+1) 
   = \Pr\big[G_2 \leq \Qsf^{-1}(\epsilon_{0}), G_2 \leq \Qsf^{-1}(\epsilon_{1}) \big]
\\&= \Pr\big[G_2 \leq \min(\Qsf^{-1}(\epsilon_{0}), \Qsf^{-1}(\epsilon_{1})) \big]
%\\&= \Pr\big[G_2 \leq \Qsf^{-1}(\max(\epsilon_{0},\epsilon_{1})) \big]
%\\&= 1- \Qsf\big( \Qsf^{-1}(\min(\epsilon_{0},\epsilon_{1}) \big)
\\&= 1-\max(\epsilon_{0},\epsilon_{1});
\label{eq:sup Fdef r=+1} 
%\end{align}
%
%\begin{align}
\\
\mathsf{F}&(\epsilon_{0},\epsilon_{1};0)  
   = \Pr\big[G_2 \leq \Qsf^{-1}(\epsilon_{0}), G_3 \leq \Qsf^{-1}(\epsilon_{1}) \big]
\\&= \Pr\big[G_2 \leq \Qsf^{-1}(\epsilon_{0}) \big]  \Pr\big[G_3 \leq \Qsf^{-1}(\epsilon_{1}) \big]
%\\&= \big(1- \Qsf\big(\Qsf^{-1}(\epsilon_{0})\big)\big) \big(1- \Qsf\big(\Qsf^{-1}(\epsilon_{1})\big)\big)
\\&=(1-\epsilon_{0})(1-\epsilon_{1});
\label{eq:sup Fdef r=0} 
%\end{align}
%
%\begin{align} 
\\
\mathsf{F}&(\epsilon_{0},\epsilon_{1};-1)
   = \Pr\big[G_2 \leq \Qsf^{-1}(\epsilon_{0}), -G_2 \leq \Qsf^{-1}(\epsilon_{1}) \big] 
\\&= \Pr\big[\Qsf^{-1}(1-\epsilon_{1})  \leq G_2 \leq \Qsf^{-1}(\epsilon_{0}) \big] 1_{\{ 1-\epsilon_{1} \geq \epsilon_{0} \}}
\\&= [1-\epsilon_{1}-\epsilon_{0}]^+.
\label{eq:sup Fdef r=-1} 
\end{align} 
\label{eq:sup Fdef vs r} 
%otherwise it must be evaluated numerically.
\end{subequations}
We thus conclude that the ``error rates region'' $\{ (\epsilon_{0},\epsilon_{1})\in[0,1]^2 : \mathsf{F}(\epsilon_{0},\epsilon_{1};r) \geq 1 - \varepsilon \}$ 
%as defined similarly to~\eqref{eq:sup globalerror} 
monotonically enlarges with $r$ from the triangle $\epsilon_{0} + \epsilon_{1} \leq \varepsilon$ for $r=-1$, to the square $\max(\epsilon_{0},\epsilon_{1})  \leq \varepsilon$ for $r=+1$, as depicted in 
Fig.~\ref{fig:epsregion}. This is the set of reliability pairs that we can optimize over in the superposition coding inner bound for receiver~1. Indeed, consider the function $1-\mathsf{F}(\epsilon_{\rm sat},\epsilon_{\rm cc};r)=\epsilon_1$, which is the average probability of error at receiver~1. It includes two terms: $\epsilon_{\rm sat}$ is related to the reliability of decoding the satellite codeword after having stripped the contribution of the cloud center; and $\epsilon_{\rm cc}$ is related to the probability of decoding in error the cloud center codeword (and thus also the satellite). Overall, the optimization in the superposition coding achievable region implies that we can choose the best reliability allocation among these two decoding steps in order to achieve an overall reliability $\epsilon_1$ at receiver~1. As this optimization concerns a single user, it is relevant to both the global and per-user reliability cases.
For global error, a further optimization step in the achievable region is possible: we can choose overall reliability $\epsilon_1$ at receiver~1 and $\epsilon_2$ at receiver~2 such that $1-(1-\epsilon_1)(1-\epsilon_2) \leq \varepsilon$, where $\varepsilon$ is the maximum global average probability of error. Therefore, we see that reliability optimization can be leveraged to optimize the performance of downlink systems with latency constraints.  
%%%{\red DO U AGREE ON INTERPRETATION?}{\magenta with simultaneous decoding I don't think we should use $\epsilon_{cc}$.  I think it should be $\epsilon_{\text{combined}}$ I guess it's effectively the same, the receiver is left with the correct sat but incorrect cloud center, but functionally it is an error in the threshold decoding of the combined message}
\hfill$\square$
\end{rem}

\begin{rem}[On Time Division with Global Error]\label{rem:TDM}
\rm 
A baseline scheme for the case of private rates only, that is, for $R_0=0$, is the second order region achieved by Time Division Multiplexing (TDM) with power control given by
\begin{subequations}
\begin{align}
&\Rcal^\text{\rm(TDM)}(n,\varepsilon) =
\hspace*{-1cm}
\bigcup_{ \substack{ 
(\tau_1,\tau_2,\epsilon_1,\epsilon_2)\in[0,1]^4, (\alpha_1,\alpha_2)\in\mathbb{R}_+^2 : \\
\tau_1 +\tau_2 \leq 1, \ \tau_1 \alpha_1 + \tau_2 \alpha_2 \leq 1 \\
(1-\epsilon_1)(1-\epsilon_2) \geq 1-\varepsilon } }
\hspace*{-1cm}
\{ (R_1,R_2)\in \mathbb{R}^2_+ : 
\\& R_1  \leq \tau_1\kappa(\tau_1 n,\alpha_1\gamma_1,\epsilon_1) + O(\ln(\tau_1 n)/n),%, \text{\rm(weak user)}
\\& R_2  \leq \tau_2\kappa(\tau_2 n,\alpha_2\gamma_2,\epsilon_2) + O(\ln(\tau_2 n)/n) %, \text{\rm(strong user)} 
\},
\label{eq:normal approximation tdm}
\end{align}
\label{eq:TDM}
\end{subequations}

\noindent where $\tau_j n$ channel uses are allocated to receiver~$j$, subject to the total time constraint $\tau_1 +\tau_2 \leq 1$;
where power $\alpha_j P$ is allocated to receiver~$j$, subject to the average power constraint $\tau_1 \alpha_1 + \tau_2 \alpha_2 \leq 1$; and 
where $\epsilon_j$ is the reliability allocated to receiver~$j$, subject to the average probability of error constraint $(1-\epsilon_1)(1-\epsilon_2) \geq 1-\varepsilon$ (as the noises are assumed to be independent). We shall plot this region in our numerical evaluations.
%%%%{\red QUESTION: what can we say about the optimal parameters in $\Rcal^\text{\rm(TDM)}(n,\varepsilon)$?}{\magenta 
Numerically we observed that $\alpha_2$ is always greater than $\alpha_1$ for points on the boundary of $\Rcal^\text{\rm(TDM)}(n,\varepsilon)$, however the optimal parameters are difficult to describe analytically as they are linked with the optimization of the time split parameters $\tau_1,\tau_2$ and of the reliabilities $\epsilon_1,\epsilon_2$.
%%%but it is not a monotonically changing allocation along the rate region boundary. At times an increase in desired rate requires a decrease in $\alpha$ to maximize the sum rate. (This is compensated by a commiserate increase in the user's $\tau$. (IE more time,less power)). In contrast increasing the desired rate for either user always requires a decrease in $\epsilon$ to maximize the sum rate. Should we mention the problems with small $n$ for TDMA?
%%%}
\hfill$\square$
\end{rem}

\begin{rem}[On Concatenate \& Code with Global Error]\label{rem:CCP}
\rm 
The choice $\beta=1$ in $\Rcal^{\text{\rm(SUP)}}(n,\varepsilon)$ means that no satellite codewords are sent, that is both users decode the same codeword with each user recovering their message from some fraction of the bits encoded. In \cite{9500658_psdtbs_icc2021} we referred to this case as {\it Concatenate \& Code Protocol} (CCP). CCP is obtained as a special case of $\Rcal^{\text{\rm(SUP)}}(n,\varepsilon)$ for $\alpha=0$ and $\epsilon_{10}=0$, resulting in $\Vsf^\prime(0, \gamma_2)=\Vsf(\gamma_2)$.
\begin{subequations}
Thus, the CCP region is
\begin{align}
&\Rcal^\text{\rm(CCP)}(n,\varepsilon) =
%\hspace*{-.5cm}
\{ (R_0,R_1,R_2)\in \mathbb{R}^3_+ : R_0+R_1+R_2 
\\&\leq 
%\hspace*{-.5cm}
\max_{\substack{ (\epsilon_1,\epsilon_2)\in[0,1]^2 \\ (1-\epsilon_2)(1-\epsilon_1)\geq 1-\varepsilon} }
\hspace*{-.5cm}
\min( \kappa(n, \gamma_1, \epsilon_1), \kappa(n, \gamma_2, \epsilon_2))
 + O_{\ln(n)/n}
\}.
\end{align}
\label{eq:CCP}
\end{subequations}
We note that it is possible to have $\kappa(n, \gamma_1, \epsilon_1) < \kappa(n, \gamma_2, \epsilon_2)$ even under the assumption $\gamma_1 > \gamma_2$ if $\epsilon_1 \ll \epsilon_2$. Numerically we observed that the optimal reliability allocation
%in $\Rcal^\text{\rm(CCP)}(n,\varepsilon)$ 
is $\epsilon_1 \leq \epsilon_2 \approxeq \varepsilon$ such that $\kappa(n, \gamma_1, \epsilon_1)  = \kappa(n, \gamma_2, \epsilon_2)$.
%{\red QUESTION: TRUE?}.{\magenta True, but $\kappa(n, \gamma_1, \epsilon_1) = \kappa(n, \gamma_2, \epsilon_2)$ and we could note that this implies a constant reliability allocation across the achievable rate region as $\lambda$ terms aren't changing.}
\hfill$\square$
\end{rem}

\begin{rem}[On Superposition Coding without Rate Splitting with Global Error]\label{rem:SUPnoRS}
\rm 
An achievable region without rate splitting is obtained by setting $\beta=0$ in $\Rcal^\text{\rm(SUP)}(n,\varepsilon)$. In this case we numerically observed that the sum-rate bound is always tight (that is, eq\eqref{eq:normal approximation sup cloudcenter}+eq\eqref{eq:normal approximation sup satellite}=eq\eqref{eq:normal approximation sup sumrate}, and that the optimal reliability allocation is such that $\epsilon_{11} \ll \epsilon_{10} \approxeq \epsilon_1$. We shall refer to this region as $\Rcal^{\text{\rm(SUPnoRS)}}(n,\varepsilon)$, given by
\begin{subequations}
\begin{align}
&\Rcal^{\text{\rm(SUPnoRS)}}(n,\varepsilon) 
=\hspace*{-.5cm}
\bigcup_{\substack{ (\alpha,\epsilon_2,\epsilon_1)\in[0,1]^3 \\
\text{ eq\eqref{eq:normal approximation sup cloudcenter}+eq\eqref{eq:normal approximation sup satellite}=eq\eqref{eq:normal approximation sup sumrate}} \\
(1-\epsilon_2)(1-\epsilon_1)\geq 1-\varepsilon 
}}
\hspace*{-.5cm}
\Big\{ (R_0,R_1,R_2)\in \mathbb{R}^3_+ :
\\& R_0+R_2 \leq \Csf\left(\frac{(1-\alpha)\gamma_2}{1+\alpha \gamma_2}\right)
\\& \quad -\sqrt{\frac{1}{n} \Vsf^\prime(\alpha \gamma_2, \gamma_2)} \Qsf^{-1}\left( \epsilon_2 \right) + O_{\ln(n)/n},
\\& R_1  \leq \Csf\left(\alpha\gamma_1\right)
-\sqrt{\frac{1}{n} \Vsf(\alpha \gamma_1)} \Qsf^{-1}\left( \epsilon_1 \right) + O_{\ln(n)/n}
\Big\}.
\end{align}
\label{eq:SUPonly} 
\end{subequations}
\hfill$\square$
\end{rem}
%%%{\red QUESTION TRUE?}{\magenta This is not true.  The sum rate is in fact always active.  If it were not then sat reliability could be loosened and combined could be tightened until it was tight, resulting in a higher sum rate at the same cloud center rate. The equation relating $\epsilon_{11},\epsilon_{10}, \epsilon_1$ is accurate}, that is, {\red DT: then the region below is not correct; why was it not corrected?}

\begin{rem}
\rm
One can define regions $\Rcal^\text{\rm(TDM)}(n,\epsilon_1,\epsilon_2)$
(akin to~\eqref{eq:TDM}),
$\Rcal^\text{\rm(CCP)}(n,\epsilon_1,\epsilon_2)$
(akin to~\eqref{eq:CCP}), and
$\Rcal^{\text{\rm(SUPnoRS)}}(n,\epsilon_1,\epsilon_2)$
(akin to~\eqref{eq:SUPonly})
for per-user error by removing the constraint $(1-\epsilon_2)(1-\epsilon_1)\geq 1-\varepsilon$ in the respective optimizations. The order of superposition can also be swapped in order to possibly obtain larger achievable regions.
\hfill$\square$
\end{rem}

\begin{figure*}%[t]
  \centering
\begin{tabular}{|c c c|}
\hline
$\gamma_1=15,\gamma_2=10$& & \\\hline
\includegraphics[align=c,width=.31\linewidth]{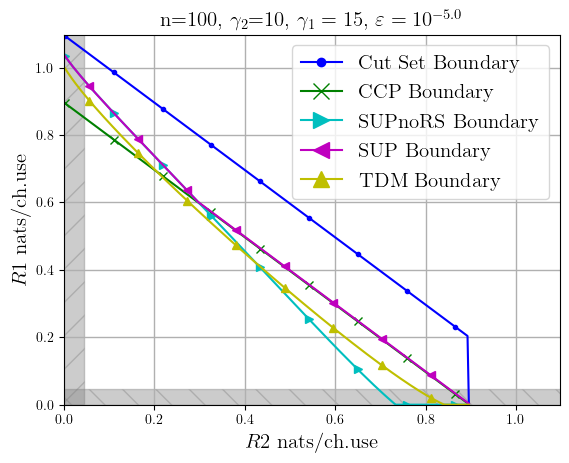} & \includegraphics[align=c,width=.31\linewidth]{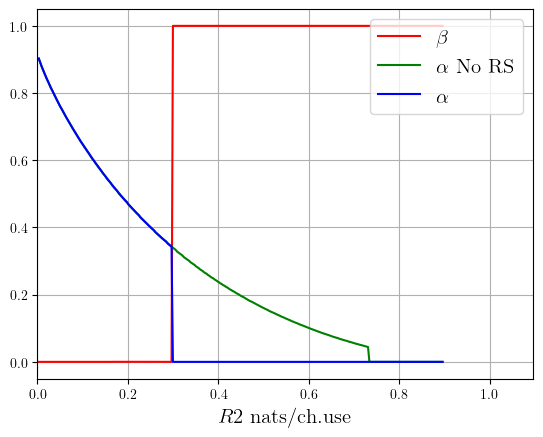} & \includegraphics[align=c,width=.31\linewidth]{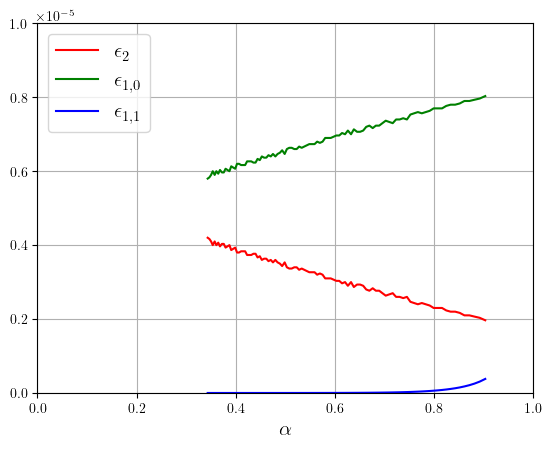}\\
\hline
Increase $\gamma_1\to40$ & &
\\\hline
\includegraphics[align=c,width=.31\linewidth]{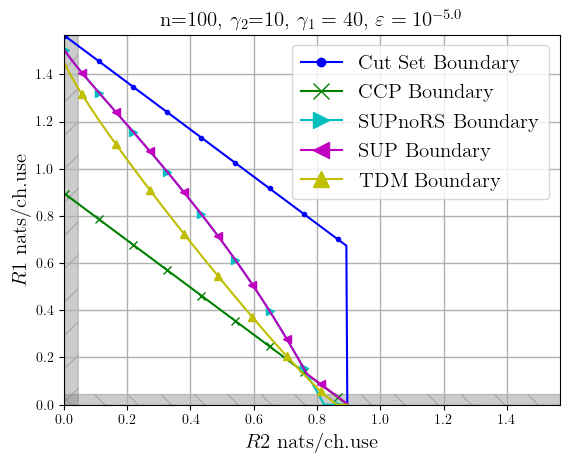} & \includegraphics[align=c,width=.31\linewidth]{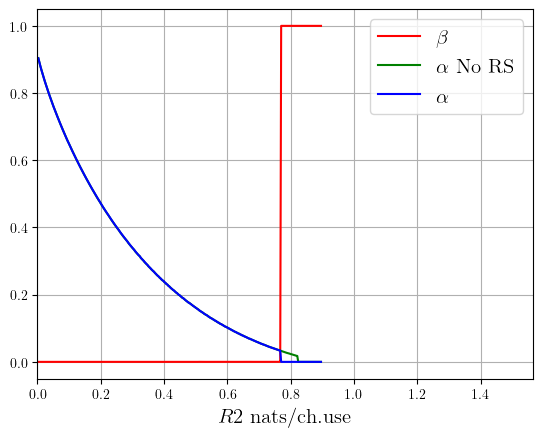} & \includegraphics[align=c,width=.31\linewidth]{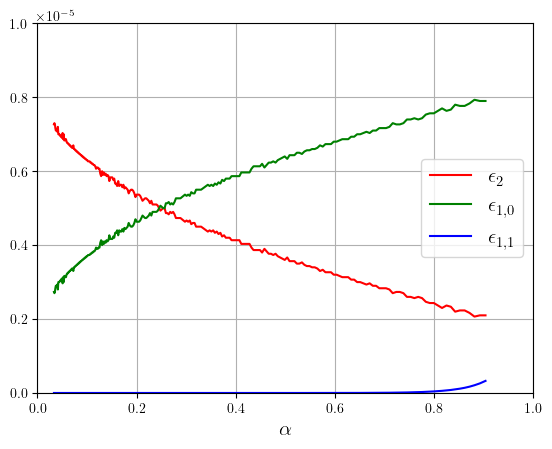}
\\ \hline
%Increase $n \to 512$
%\\ \hline
%\includegraphics[align=c,width=.31\linewidth]{./awgn-bc-plots-2022/rateRegion/1global_10.0_40.0_1e-05_512.png} & %\includegraphics[align=c,width=.31\linewidth]{./awgn-bc-plots-2022/alphaDelta/2global_10.0_40.0_1e-05_512} & %\includegraphics[align=c,width=.31\linewidth]{./awgn-bc-plots-2022/epsilonVsAlpha/3bglobal_10.0_40.0_1e-05_512}
%\\\hline
\end{tabular}
\caption{Left: Achievable rate regions Center: Power and rate split allocations as a function of the rate allocated to user 2. Right: Reliability allocations as function of the optimal power split.}
\label{fig:numResults}
\end{figure*}

\section{Numerical Evaluations}
\label{sec:numericalEvaluations}
%{\red DT: I removed some figures because we had too many pages. In case I removed figures that were introduced for a specific purpose please feel free to insert instead of some other figures. Also make sure to spell out walk the figure is showing.}

%\subsection{Global Error Constraint}
We start by giving numerical evaluations of the second order rate region in Theorem~\ref{thm:main} for private rates only, that is, $R_0=0$. 
%\nbl{ND: Why no numerical with common rates at all?} DT: add it to Rweak
Numerically we observed that:
(i) achievable regions are not convex when the normal approximation terms become comparable to $\ln(n)/n$, which is the areas highlighted in grey in the figures; 
%%%{\red NOT EVEN TDM, RIGHT? If so, WHY? HOW TO CONVEXIFY IT?}{\magenta TDM is not convex near the axes because the low number of channel uses assigned to the user near a rate of zero means the finite blocklength penalty is large and potentially inaccurate. It is close to being convex inside of the grey bars. You can convexify from points to the right via CCP, but this does not help with the SUP. We could come up with a convexification via TDM but it would involve increasing the block size.  (IE find the midpoint between two $n=100$ points would require $n=200$), possibly worth more analysis}; 
and
(ii) $\beta\in(0,1)$ never gives a point on the boundary of the region $\Rcal^{\text{\rm(SUP)}}(n,\varepsilon)$, that is, $\Rcal^{\text{\rm(SUP)}}(n,\varepsilon)$ is the union of $\Rcal^{\text{\rm(SUPnoRS)}}(n,\varepsilon)$ in~\eqref{eq:SUPonly} and $\Rcal^{\text{\rm(CCP)}}(n,\varepsilon)$ in~\eqref{eq:CCP}.

In Fig.~\ref{fig:numResults} %and~\ref{fig:numResults2}
we plot in the left column the region $\Rcal^{\text{\rm(SUPnoRS)}}(n,\varepsilon)$ in~\eqref{eq:SUPonly} %(labeled "SUPnoRS") 
and $\Rcal^{\text{\rm(CCP)}}(n,\varepsilon)$ %(labeled "CCP")
 in~\eqref{eq:CCP}. As a baseline, we plot $\Rcal^{\text{\rm(TDM)}}(n,\varepsilon)$ in~\eqref{eq:TDM}. As a converse bound, we plot $\Rcal^{\text{\rm(CS)}}(n,\varepsilon)$ in~\eqref{eq:cutset globalerror}. In all plots, we set $\gamma_2=10$, and $\varepsilon=10^{-5}$. 
We neglect the third-order term $O_{\ln(n)/n}$. 
%The ``gray bands'' in the left column plots in Fig.~\ref{fig:numResults} and~\ref{fig:numResults2} represents rates that are of the order of $\ln(n)/n$ for which the third-order term is no longer negligible. 
We note that when the SNRs are comparable and $n$ is not too large, %%%{\red RIGHT?} {\magenta yes},
 CCP is superior to SUPnoRS when the user with the largest SNR has a relatively low rate. %%%{\red RIGHT?} {\magenta correct}. 
In the second column in Fig.~\ref{fig:numResults} %and~\ref{fig:numResults2} 
we show the optimal power and rate split vs $R_2$. We observe the sharp transition in $\beta$ that marks when CCP outperforms SUPnoRs. Improved channel conditions of the strong user decrease the $\alpha$ at which this transition occurs. As the SNRs become more dissimilar, the portion of the achievable rate region boundary attained by CCP decreases. 
The right column in Fig.~\ref{fig:numResults} %and~\ref{fig:numResults2} 
shows the optimal reliability allocation vs $\alpha$. We observe that the $\epsilon_{1,1}$ term indicates that in the optimal allocation the strong user recovers the cloud center with a very high reliability across the rate region. More generally, we see that a relaxation of reliability for a user recovering their message is optimal as the rate demands of that user increase.

%\begin{align}
%\left( 
%%\beta,\frac{\epsilon_{10}}{\varepsilon}, \frac{\epsilon_{11}}{\varepsilon},
%\frac{1-\mathsf{F}(\epsilon_{10},\epsilon_{11};r(\alpha\gamma_1, \gamma_1))}{\varepsilon},
%\frac{\epsilon_2}{\varepsilon},
%%
%\frac{R_2}{\kappa(n,\gamma_2,\varepsilon),} 
%\frac{R_1}{\kappa(n,\gamma_1,\varepsilon)}
%\right) \in[0,1]^4
%\end{align}
%vs $\alpha$ that attain the points on the boundary of SUPnoRS achievable region. {\red TO DO; WHAT DO WE NOTICE? Should we also plot the same for tdm and ccp? Add grey vertical bars for those alphas that give rate values too low.}

\begin{figure}%[t]
\centering
\includegraphics[width=.86\columnwidth]{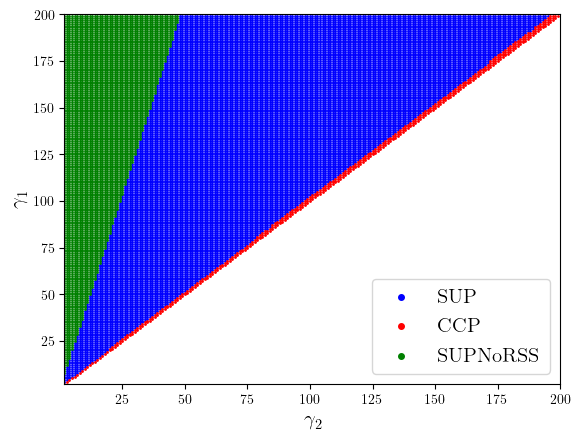} 
\caption{`Simplest` coding scheme required to obtain the largest achievable rate region for two users with varying channel conditions and a constant $\varepsilon=0.1$ and n = 100.}
\label{fig:codingVsGamma}
\end{figure}
%%%\begin{figure*}[t]
%%%\centering
%%%\begin{tabular}{|c | c | c | c| c|}
%%%\hline
%%%&$\varepsilon=10^{-1}$&$\varepsilon=10^{-3}$&$\varepsilon=10^{-5}$\\\hline
%%%n=100& \includegraphics[align=c,width=.3\linewidth]{awgn-bc-plots-2022/OptimalCoding/epsilon_0.1_n100.png} & \includegraphics[align=c,width=.3\linewidth]{awgn-bc-plots-2022/OptimalCoding/epsilon_0.001_n100.png} & \includegraphics[align=c,width=.3\linewidth]{awgn-bc-plots-2022/OptimalCoding/epsilon_1e-05_n100.png}\\ \hline
%%%n=300& \includegraphics[align=c,width=.3\linewidth]{awgn-bc-plots-2022/OptimalCoding/epsilon_0.1_n300.png} & \includegraphics[align=c,width=.3\linewidth]{awgn-bc-plots-2022/OptimalCoding/epsilon_0.001_n300.png} & \includegraphics[align=c,width=.3\linewidth]{awgn-bc-plots-2022/OptimalCoding/epsilon_1e-05_n300.png}\\\hline 
%%%n=500& \includegraphics[align=c,width=.3\linewidth]{awgn-bc-plots-2022/OptimalCoding/epsilon_0.1_n500.png} & \includegraphics[align=c,width=.3\linewidth]{awgn-bc-plots-2022/OptimalCoding/epsilon_0.001_n500.png} & \includegraphics[align=c,width=.3\linewidth]{awgn-bc-plots-2022/OptimalCoding/epsilon_1e-05_n500.png}\\\hline
%%%\end{tabular}
%%%\caption{Suitable Coding Schemes. {\red just one figure, as they are all the same! and we need space.}}
%%%\label{fig:codingVsGamma}
%%%\end{figure*}

In Fig.~\ref{fig:codingVsGamma} we present a plot showing the coding scheme used to achieve the largest achievable regions across a set of channel conditions $\gamma_1,\gamma_2\in[2,50]$ for %a global reliability requirement 
$\varepsilon=10^{-1}$ and a blocklength $n=100$. For each point in the plot, %the specified achievable rate region was 
we evaluated the CCP, SUP, and SUPNoRS regions. The points are colored based on the ``simplest'' coding scheme that achieves the largest achievable region for meaningful rates, that is, larger than $\ln(n)/n$, for both users. Here ``simplicity'' is a somewhat arbitrary measure we define as $\{$CCP, SUPNoRS, SUP$\}$ with complexity increasing from left to right. This intuitively corresponds to the complexity of the coding scheme implementation by broadcaster and receiver, but more importantly we use it to illustrate the fact that for a very large set of channel conditions and reliability requirements, rate splitting (either as part of SUP or alone as CCP) is required to achieve the largest achievable rate regions. In the global reliability case this plot is symmetric about the line $\gamma_1=\gamma_2$ so only the top half is plotted Increasing the global reliability requirement increases the size of this set while increasing  the blocklength reduces it. In effect, as might be expected, increasing the blocklength or decreasing reliability requirements makes the second order region more and more similar to the (infinite blocklength) capacity region.

\begin{figure*}%[t]
\centering
\begin{tabular}{|c|c |c|}
\hline
$\gamma_1=35,\gamma_2=30$
\\\hline
&$\varepsilon_1=10^{-5},\varepsilon_2=0.1$&$\varepsilon_1=0.1,\varepsilon_2=10^{-5}$
\\\hline
n=100&\includegraphics[align=c,width=.3\linewidth]{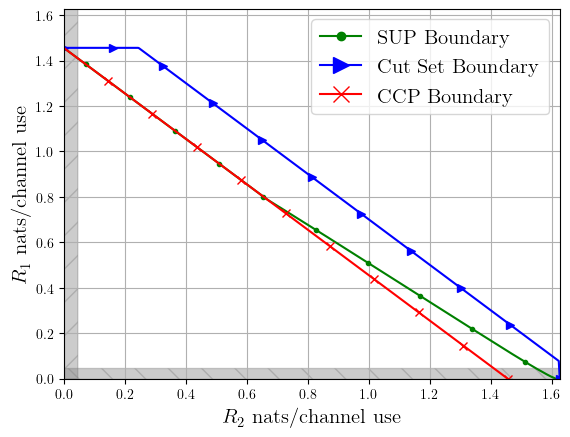}
&\includegraphics[align=c,width=.3\linewidth]{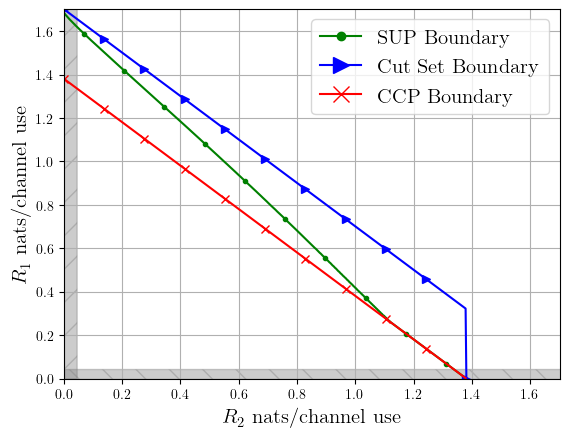}
\\\hline
n=5000&\includegraphics[align=c,width=.3\linewidth]{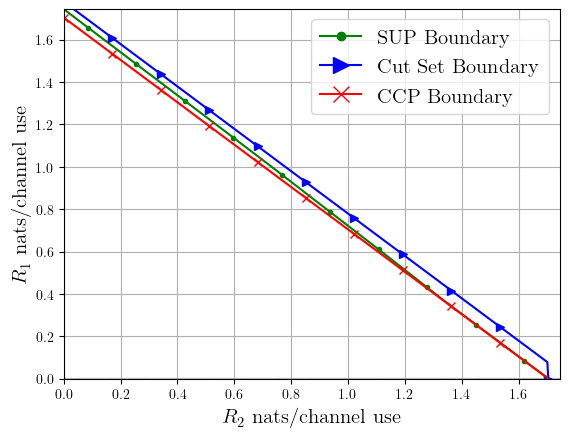} 
& \includegraphics[align=c,width=.3\linewidth]{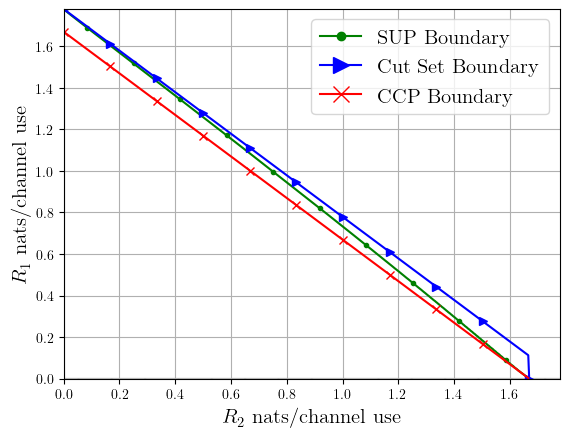} 
\\\hline
\end{tabular}
\caption{\small Per-user error constraint achievable rate regions.}
\label{fig:numResultsPerUser}
\end{figure*}
 %In Fig.~\ref{fig:numResults} and~\ref{fig:numResults2} we plot in the left column the region $\Rcal^{\text{\rm(SUPnoRS)}}(n,\varepsilon)$ in~\eqref{eq:SUPonly} %(labeled "SUPnoRS") 
%and $\Rcal^{\text{\rm(CCP)}}(n,\varepsilon)$ %(labeled "CCP")
 %in~\eqref{eq:CCP}. As a baseline scheme, we plot $\Rcal^{\text{\rm(TDM)}}(n,\varepsilon)$ in~\eqref{eq:TDM}. As a converse bound, we plot $\Rcal^{\text{\rm(CS)}}(n,\varepsilon)$ 

%\subsection{Per-User Error Constraint}
%\label{perUserNumericalSection}
We now show plots for the per-user error requirements.
In Fig.~\ref{fig:numResultsPerUser} we present $\Rcal^{\text{\rm(CCP)}}(n,\epsilon_1,\epsilon_2)$, $\Rcal^{\text{\rm(SUP)}}(n,\epsilon_1,\epsilon_2)$, and $\Rcal^{\text{\rm(CS)}}(n,\epsilon_1,\epsilon_2)$ for four scenarios.  In all scenarios the SNRs are $\gamma_1=35$ and $\gamma_2=30$. The scenario's blocklength and reliability requirements are varied. For the top row $n=100$, and for the bottom row $n=5000$. The reliability constraints are varied from left to right.  On the left, user 2 has a more relaxed reliability requirement of $0.9\%$ and user 1 has a high reliability requirement of $99.999\%$. When user 2 has a larger  point-to-point second order capacity (top left), a larger achievable rate region is found by encoding user 1's message in the cloud center.  When $n$ is increased to $5000$, user 2 no longer has a larger point-to-point second order capacity and the capacity-achieving superposition ordering provides the largest achievable rate region. On the right, the plots maintain a similar shape as $n$ is increased as the point-to-point second order capacity ordering does not change.

\begin{figure}%[t]
\centering
\includegraphics[width=.86\columnwidth]{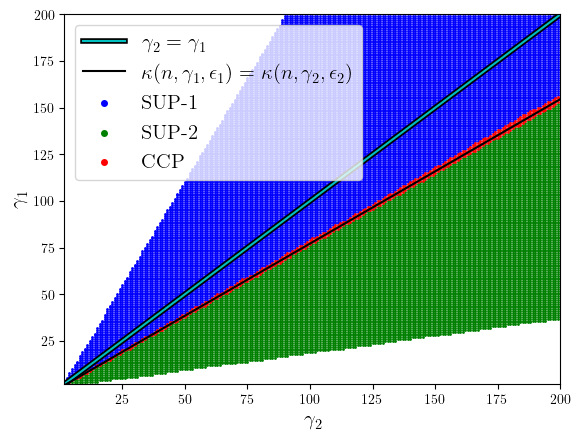} 
\caption{\small `Simplest` coding scheme required to obtain the largest achievable rate region for two users with varying channel conditions and a constant $\varepsilon_1=0.1$, $\varepsilon_2=0.001$ and $n = 100$.}
\label{fig:suitableCodingsPerUser}
\end{figure}

In Fig.~\ref{fig:suitableCodingsPerUser} we present (as in Fig.~\ref{fig:codingVsGamma}) the coding schemes that achieve the largest achievable second order rate region for thousands of combinations of channel conditions.  In each case user 2 has a higher reliability requirement of $99.999\%$ while the reliability of user 1 is $90\%$. The blocklength is fixed $n=100$.
Points marked as SUP-1 are channel conditions and reliability requirements where encoding user 2's message in the cloud center gives the largest region.  Points marked SUP-2 are channel conditions where encoding user 1's message in the cloud center gives the largest region.  

Points marked as CCP are channel conditions in which neither SUP ordering produces points on the achievable rate region boundary beyond what is produced by CCP.  The achievable rate region formed by either SUP-1 or SUP-2 consists only of points where $\beta=1$. This band clusters around and includes the line where the P2P second order capacities are equal.  

Finally, unmarked points correspond to channel conditions in which rate splitting is not required to achieve the largest region for any rate larger than $\ln(n)/n$. In these cases, a standard capacity achieving superposition code scheme achieves the best known finite blocklength achievable rate region.

%%%\begin{figure*}[t]
%%% \centering
%%%\begin{tabular}{|c| c |c|}
%%%\hline
%%%&$\varepsilon_1=10^{-1},\varepsilon_2=10^{-3}$&$\varepsilon_1=10^{-1},\varepsilon_2=10^{-5}$\\
%%%\hline
%%%n=100& \includegraphics[align=c,width=.45\linewidth]{awgn-bc-plots-2022/perUser/epsilon1_0.1_epsilon2_0.001_n100.0.png} & \includegraphics[align=c,width=.45\linewidth]{awgn-bc-plots-2022/perUser/epsilon1_0.1_epsilon2_1e-05_n100.0.png}
%%%\\
%%%\hline
%%%n=500& \includegraphics[align=c,width=.45\linewidth]{awgn-bc-plots-2022/perUser/epsilon1_0.1_epsilon2_0.001_n500.0.png} & \includegraphics[align=c,width=.45\linewidth]{awgn-bc-plots-2022/perUser/epsilon1_0.1_epsilon2_1e-05_n500.0.png}
%%%\\\hline
%%%\end{tabular}
%%%\caption{Per-User Error Suitable Coding Schemes. {\red just one figure, as they are all the same! and we need space.}}
%%%\label{fig:suitableCodingsPerUser}
%%%\end{figure*}

\section{Converse Bound Proof}
\label{sec:converse}
%For the sake of a simpler notation, 
We shall set $R_0=0$ at the beginning of this section in order to simply the notation. 
We shall also omit to explicitly write the event $\{ (W_1,W_2)~\text{sent}\}$ within the probabilities of error.
For the two-user AWGN BC with global error bounded bounded by $\varepsilon$, 
%where the transmitted messages are $(W_1,W_2)$,
we trivially have 
\begin{subequations}
\begin{align} 
  1-\varepsilon 
 \leq &\Pr[\mathsf{dec}_1(Y_1^n)=W_1 \cap \mathsf{dec}_2(Y_2^n)=W_2] %two separate decoders
\\ \leq \min\{
     &\Pr[\mathsf{dec}_\text{genie}(Y_1^n,Y_2^n)=(W_1,W_2)], %a decoder that has both outputs
\label{eq:cutset1 both}
\\   &\Pr[\mathsf{dec}_1(Y_1^n)=W_1], %forget about RX2
\label{eq:cutset1 1}
\\   &\Pr[\mathsf{dec}_2(Y_2^n)=W_2]  %forget about RX1
\label{eq:cutset1 2}
     \},
\end{align}
\label{eq:cutset1}
where each of the terms in the minimum function in~\eqref{eq:cutset1} relates to the performance of a Gaussian point-to-point channel.
In particular, the probability of correct decoding in~\eqref{eq:cutset1 both} is that of a gene-aided receiver that has both channel outputs,
and those in~\eqref{eq:cutset1 1} and~\eqref{eq:cutset1 2} correspond to considering the requirement for one of the users only.
\end{subequations}
Therefore, %by using the same argument that leads to~\eqref{eq:normal approximation M*}, 
an outer bound for $\Ccal(n,\varepsilon)$ from~\eqref{eq:cutset1} is %(note same $\epsilon$ everywhere!)
\begin{subequations}
\begin{align} 
\Ccal(n,\varepsilon) &\subseteq
\Big\{ (R_1,R_2)\in \mathbb{R}^2_+ :
\\ R_1    &\leq \kappa(n,\gamma_1,\varepsilon) + O_{\ln(n)/n},
\\ R_2    &\leq \kappa(n,\gamma_2,\varepsilon) + O_{\ln(n)/n},
\\ R_1+R_2&\leq \kappa_\text{SIMO}(n,\gamma_1,\gamma_2,\varepsilon) + O_{\ln(n)/n}
\label{eq:cutset try1 sumrate}
\Big\},
\end{align}
\label{eq:cutset try1}
where $\kappa_\text{SIMO}(n,\gamma_1,\gamma_2,\varepsilon)$ in~\eqref{eq:cutset try1 sumrate} is the second order normal approximation for the Gaussian point-to-point SIMO channel (with SNRs at the two receive antennas given by $\gamma_1$ and $\gamma_2$) with error rate $\varepsilon$; this bound depends on the correlation on the noises on the two antennas.
%\nbl{[DONEbySOMEONE?].} \blue{PS: I believe this is a specialization of [Theorem 3] in "Coherent Multiple-Antenna Block-Fading Channels at Finite Blocklength"}
\end{subequations}
In~\cite{9500658_psdtbs_icc2021} we wrote that the sum-rate in~\eqref{eq:cutset try1 sumrate} can be replaced by 
\begin{align} 
R_1+R_2 &\leq \kappa(n,\max(\gamma_1,\gamma_2),\varepsilon) + O_{\ln(n)/n},
\label{eq:cutset try2}
\end{align} 
which is true only for the physically degraded BC;
%\footnote{Indeed, WLOG let $\sigma_1 \leq \sigma_2$, for $X \to Y_1 \to Y_2$, let so that 
in this case $Y_2 = Y_1+Z_0$ with $Z_0\sim \Ncal(0,\sigma_2^2-\sigma_1^2)$ independent of $Z_1$, and thus $\mathsf{dec}_\text{genie}(Y_1^n,Y_2^n) = \mathsf{dec}(Y_1^n)$, but for the general case we cannot draw the same conclusion. %use~\eqref{eq:sumrate phy deg}.
%\begin{subequations}
%\begin{align} 
%  1-\varepsilon 
%  &\leq \Pr[\mathsf{dec}_\text{genie}(Y_1^n,Y_2^n)=(W_1,W_2)]_{Y_2=Y_1+Z_0}
%\\&= \Pr[\mathsf{dec}_\text{genie}(Y_1^n,Y_1^n+Z_0^n)=(W_1,W_2)]
%\\&= \Pr[\mathsf{dec}(Y_1^n)=(W_1,W_2)].
%\end{align} 
%\label{eq:sumrate phy deg}
%\end{subequations}
%However, we do not know how in general 
%\begin{align} 
%\Pr[\mathsf{dec}_1(Y_1^n)=W_1 \cap \mathsf{dec}_2(Y_2^n)=W_2]
%\end{align} 
%comapres to
%\begin{align} 
%\Pr[\mathsf{dec}_1(Y_1^n)=W_1 \cap \mathsf{dec}_2(Y_1^n+Z_0^n)=W_2]=\Pr[\mathsf{dec}(Y_1^n)=(W_1,W_2)].
%\end{align} 
%%%I think the following should be true:
%%%\begin{align} 
%%%  &\min_{Y_2 \sim Y_1+Z_0} \Pr[\mathsf{dec}_1(Y_1^n)=W_1 \cap \mathsf{dec}_2(Y_2^n)=W_2]
%%%\\&=\Pr[\mathsf{dec}_1(Y_1^n)=W_1 \cap \mathsf{dec}_2(Y_1^n+Z_0^n)=W_2] 
%%%\\&=\Pr[\mathsf{dec}(Y_1^n)=(W_1,W_2)].
%%%\end{align} 
%
%
%==========================================================================================
%==========================================================================================
%==========================================================================================
\begin{figure*} 
\begin{tcolorbox}[colback=white]
\begin{subequations} 
Let $E_1 =\{ \mathsf{dec}_1(Y_1^n)\not=W_1 \}$ and $E_2 =\{ \mathsf{dec}_2(Y_2^n)\not=W_2 \}$ be the error events at the receivers. For $\gamma_1\geq\gamma_2$ we have
%My understanding of the argument in [https://arxiv.org/abs/2202.02110] is as follows:
\begin{align}  
%\Ccal(n,\varepsilon) 
%  &= \cup_{\mathsf{enc},\mathsf{dec}_1,\mathsf{dec}_2} \{ (R_1,R_2):  \Pr[E_1 \cup E_2] \leq \varepsilon \}
%\\&= \cup_{\mathsf{enc},\mathsf{dec}_1,\mathsf{dec}_2} \{ (R_1,R_2):  \Pr[E_2] + \Pr[E_1\setminus E_2] \leq \varepsilon \}
%\\&\subseteq \cup_{\mathsf{enc},\mathsf{dec}_1,\mathsf{dec}_2} \{ (R_1,R_2):  \Pr[E_2] \leq \varepsilon \}
%\\&= \cup_{\mathsf{enc},\mathsf{dec}_2} \{ R_2:  \Pr[\mathsf{dec}_2(Y_2^n)\not=W_2] \leq \varepsilon \};
%\\
\Ccal(n,\varepsilon) 
  &= \cup_{\mathsf{enc},\mathsf{dec}_1,\mathsf{dec}_2} \{ (R_1,R_2):  \Pr[E_1 \cup E_2] \leq \epsilon \}
\\&= \cup_{\mathsf{enc},\mathsf{dec}_1,\mathsf{dec}_2} \{ (R_1,R_2):  \Pr[E_1] + \Pr[E_2\setminus E_1] + \Pr[E_2] + \Pr[E_1\setminus E_2] \leq 2\varepsilon \}
\\&\subseteq \cup_{\mathsf{enc},\mathsf{dec}_1,\mathsf{dec}_2} \{ (R_1,R_2):  \Pr[E_1] + \Pr[E_2] \leq 2\epsilon \}
\\&= \cup_{\mathsf{enc},\mathsf{dec}_1,\mathsf{dec}_2} \{ (R_1,R_2):  \Pr[\mathsf{dec}_1(Y_1^n)\not=W_1] + \Pr[\mathsf{dec}_2(Y_1^n+Z_0^n)\not=W_2] \leq 2\varepsilon \} 
\label{eq:cutset2 only marginals matter} %\ \text{\rm(only marginals matter)}
\\&\subseteq  \cup_{\mathsf{enc},\mathsf{dec}_1,\mathsf{dec}_2} \{ (R_1,R_2):  \Pr[\mathsf{dec}_1(Y_1^n)\not=W_1] + \Pr[\mathsf{dec}_2(Y_1^n)\not=W_2] \leq 2\varepsilon \} 
\label{eq:cutset2 monotonicity in SNR} %\ \text{\rm(monotonicity in SNR)}
\\&\subseteq  \cup_{\mathsf{enc},\mathsf{dec}_0} \{ (R_1,R_2):  \Pr[\mathsf{dec}_0(Y_1^n)\not=(W_1,W_2)] \leq 2\varepsilon \},
\end{align}
\label{eq:cutset2} 
where in~\eqref{eq:cutset2 only marginals matter} %only the marginal distributions matter, i.e., 
we used $Y_2^n \sim Y_1^n+Z_0^n$, and in~\eqref{eq:cutset2 monotonicity in SNR} the monotonicity in SNR.
\end{subequations} 
\end{tcolorbox}
\end{figure*}
%Say (1,1) sent
%%\begin{align*}
%%  &  \Pr[\mathsf{dec}_1(Y_1^n)\not=1] + \Pr[\mathsf{dec}_2(Y_1^n)\not=1]
%%\\&=2\Pr[\mathsf{dec}_1(Y_1^n)\not=1, \mathsf{dec}_2(Y_1^n)\not=1] 
%%\\&+ \Pr[\mathsf{dec}_1(Y_1^n)\not=1, \mathsf{dec}_2(Y_1^n)=1] 
%%\\&+ \Pr[\mathsf{dec}_1(Y_1^n)=1,\mathsf{dec}_2(Y_1^n)\not=1]
%%\\&\geq \Pr[\mathsf{dec}_1(Y_1^n)\not=1, \mathsf{dec}_2(Y_1^n)\not=1] 
%%\\&+ \Pr[\mathsf{dec}_1(Y_1^n)\not=1, \mathsf{dec}_2(Y_1^n)=1] 
%%\\&+ \Pr[\mathsf{dec}_1(Y_1^n)=1,\mathsf{dec}_2(Y_1^n)\not=1]
%%\\&=\Pr[\mathsf{dec}_0(Y_1^n)\not=(1,1)].
%%\end{align*} 
%\begin{align*}
%  &  \Pr[\hat{W}_1\not=1] + \Pr[\hat{W}_2\not=1]
%\\&=2\Pr[\hat{W}_1\not=1, \hat{W}_2\not=1] 
%  + \Pr[\hat{W}_1\not=1, \hat{W}_2=1] 
%  + \Pr[\hat{W}_1=1,\hat{W}_2\not=1]
%\\&\geq \Pr[\hat{W}_1\not=1, \hat{W}_2\not=1] 
%   + \Pr[\hat{W}_1\not=1, \hat{W}_2=1] 
%   + \Pr[\hat{W}_1=1,\hat{W}_2\not=1]
%\\&=\Pr[\widehat{(W_1,W_2)}\not=(1,1)].
%\end{align*} 
%==========================================================================================
%==========================================================================================
%==========================================================================================
%
%
Next, we provide a derivation of~\cite[Corollary 1]{https://doi.org/10.48550/arxiv.2202.02110} that generalizes straightforwardly to any number of users.
From the series of inclusions in~\eqref{eq:cutset2} at the top of the next page, we can bound the sum-rate for the case of arbitrarily correlated noises as
\begin{align}
R_1+R_2&\leq \kappa(n,\max(\gamma_1,\gamma_2),2\varepsilon) + O_{\ln(n)/n}.
\label{eq:cutset try3}
\end{align}
Notice that the error term in~\eqref{eq:cutset try3} is $2\varepsilon$, while in~\eqref{eq:cutset try2} it was $\varepsilon$.
%Recall that~\eqref{eq:cutset try2} is only valid for physically degraded channels.
The sum-rate bound in~\eqref{eq:cutset try3} with the single-rate bounds in~\eqref{eq:cutset try1} proves the right hand side inclusion in Theorem~\ref{thm:main}, after including the common rate $R_0$ back in each bound.

\subsection{Extension to \texorpdfstring{$K$}{K} users}\label{sec:outK}
The reasoning in~\eqref{eq:cutset2} extends to the case of $K$ users and gives, in the case of private rates only, the bound
\begin{subequations}
\begin{align}
&\Ccal(n,\varepsilon) \subseteq 
   \Big\{ (R_1,R_2,\ldots,R_K)\in \mathbb{R}^K_+ :  \forall S \subseteq [K]
\\&\sum_{j\in S} R_j \leq %n C({\max\{ \gamma_j : j\in S\}}) - \sqrt{n V({\max\{ \gamma_j : j\in S\}})}\Qsf^{-1}(|S|\epsilon)
\kappa\big(n, {\max\{ \gamma_j : j\in S\}}, |S|\epsilon\big) + O_{\ln(n)/n}
\label{eq:cutset try4:sum}
\Big\}.
\end{align}
\label{eq:cutset try4}
\end{subequations}
With common rates, the sum ``$\sum_{j\in S}R_j$'' in~\eqref{eq:cutset try4:sum} must be extended so as to include the rates of all the messages intended for the users indexed by the set $S$.

\section{Achievable Bound Proof}
\label{sec:achievability}

%The capacity region  (infinite block-length and vanishing error)  of the two-user AWGN BC is attained by superposition coding~\cite{book:ElGamal-Kim}. 
Superposition coding with rate splitting is capacity achieving for the more capable BC (and thus also for the stochastically degraded AWGN BC), and achieves~\cite[Sec 8.1]{ElGamalKim:book}
\begin{subequations}
\begin{align}
\Ccal
= \bigcup_{(\alpha,\beta)\in[0,1]^2 }
\Big\{ (R_0,R_1,R_2)&\in \mathbb{R}^3_+ :
\\ R_0+R_2+\beta R_1 &\leq \Csf\left(\frac{(1-\alpha)\gamma_2}{1+\alpha \gamma_2}\right), 
   %\Csf\left(\gamma_2\right)-\Csf\left(\alpha \gamma_2\right), %\ln\left(\frac{1+\gamma_2}{1+\alpha \gamma_2}\right)
\\ (1-\beta)R_1 &\leq \Csf\left(\alpha \gamma_1\right), %+\Csf\left(\frac{(1-\alpha)\gamma_2}{1+\alpha \gamma_2}\right),
\\ R_0+R_1+R_2  &\leq \Csf\left(\gamma_1\right)
\label{eq:2BCAWGNSup+RS last}
\Big\},
\end{align}
\label{eq:2BCAWGNSup+RS}
where $\alpha$ is the power split and $\beta$ is the rate split.
The constraint in~\eqref{eq:2BCAWGNSup+RS last} is always redundant 
%as far as capacity is concerned for
when $\gamma_1 \geq \gamma_2$, thus, the region in~\eqref{eq:2BCAWGNSup+RS} is equivalent to~\eqref{eq:2BCAWGNCapacity}, and $\beta=0$ is always optimal.
We aim to derive second order terms for%the region in
~\eqref{eq:2BCAWGNSup+RS}. % but at finite blocklength and fixed global error.
\end{subequations}

\paragraph{Rate Splitting}\label{par:ratesplit}
The message for user~1 is split as 
%\begin{align}
$m_1=(m_{10},m_{11}), \  \forall m_1 \in [M_1],$ % \quad \text{\rm(message split for strong user)}.
%\label{eq:rate split}
%\end{align}
here $m_{1j} \in [M_{1j}], j\in\{0,1\}$ and $M_{10}M_{11}=M_1$.
We construct a superposition coding scheme where the ``cloud center'' carries $m_2^\prime := (m_0,m_2,m_{10})$ and the ``satellite'' $m_1^\prime := m_{11}$;
%\begin{align}
%m_2^\prime := (m_0,m_2,m_{10}), 
%\quad \text{\rm(common message to both receivers)},
\label{eq:cloud center}
%\end{align}
%and carries
%\begin{align}
%m_1^\prime := m_{11}, 
%\quad \text{\rm(private message for the strong receiver)}.
%\label{eq:satellite}
%\end{align}
and where receiver~2 decodes the cloud center only, while receiver~1 decodes both. %the cloud center and the satellite.
Our code construction is the same as~\cite{MACDMS} for the MAC with degraded message sets: receiver~1 is exactly the same as the receiver in~\cite{MACDMS}, but in addition we must consider the decoding constraint of receiver~2 that only decodes the cloud center while treating the satellite codeword as noise. In addition we also need to include the power constraint at the transmitter. The details of the scheme are presented next.

\paragraph{Random Code Construction on the Power Sphere}\label{par:codebook}
For a power constraint $P>0$, fix real numbers $(\rho,P_1,P_2)\in[-1,1]\times \mathbb{R}_+\times \mathbb{R}_+$ such that\footnote{\label{foot:costruction2} As our proof will show later on, it suffices to consider $\rho=0$ in the following. This is so because geometrically~\cite{MACDMS} we can write the pair $(\bm{x}_1,\bm{x}_2)$ in $\mathcal{D}_n(\rho,P_1,P_2)$ in~\eqref{eq:codebook2} as 
%\begin{align}
%\bm{x}_1 = \rho \sqrt{P_1/P_2} \ \bm{x}_2 + \bm{a}_1 : 
%\begin{cases}
%\bm{x}_2 \in \mathcal{S}_{n-1}(\sqrt{n P_2}) \\
%\bm{a}_1 \in \mathcal{S}_{n-2}(\sqrt{n(1-\rho^2)P_1}) \\
%\langle \bm{a}_1 , \bm{x}_2 \rangle = 0\\
%\end{cases},
%\label{eq:costruction1}
%\end{align}
%and thus equivalently, %given $ \bm{x} = \bm{x}_1 + \bm{x}_2$ in~\eqref{eq:codebook2} and the definitions in~\eqref{eq:powalpha},
% we can write the transmitted codeword $\bm{x}$ as
\begin{align*}
\bm{x} %= \bm{x}_1 + \bm{x}_2 
=  \bm{a}_2 + \bm{a}_1: 
\begin{cases}
\bm{a}_2 \in \mathcal{S}_{n-1} (\sqrt{n \, (1-\alpha) P}) \\
\bm{a}_1 \in \mathcal{S}_{n-2}(\sqrt{n \, \alpha P}) \\
\langle \bm{a}_1 , \bm{a}_2 \rangle = 0\\
\end{cases}.
%\label{eq:costruction2}
\end{align*}
We decided to describe the scheme with any $\rho\in[-1,1]$ to make the code construction, and thus its analysis, to be essentially the same as in~\cite{MACDMS}.
}
\begin{align}
(1-\rho^2)P_1 +( \sqrt{P_2} + \rho \sqrt{P_1} )^2 = P.
\label{eq:powparams}
\end{align}
We further parameterize~\eqref{eq:powparams} as follows, for some $\alpha\in[0,1]$,
\begin{subequations}
\begin{align}
  & (1-\rho^2)P_1 = \alpha P, 
\\& ( \sqrt{P_2} + \rho \sqrt{P_1} )^2 = \xi^2 P_2  = (1-\alpha)P,
\\&\xi := 1 + \rho \sqrt{P_1/P_2}.
\label{eq:powalpha:xi}
\end{align}
\label{eq:powalpha}
In order to write~\eqref{eq:powalpha:xi} we implicitly assumed $P_2>0$, or equivalently $\alpha\not=1$; 
the extreme cases $\alpha=0$ and $\alpha=1$ will be analyzed separately in the following.
\end{subequations}
The  codebook is composed of triplets $(\bm{x}_1,\bm{x}_2,\bm{x})\in \mathbb{R}^{3n}$ from the set %$\mathcal{D}_n(\rho,P_1,P_2)$ defined as
\begin{subequations}
\begin{align}
\mathcal{D}_n(\rho,P_1,P_2) := \Big\{ 
  &(\bm{x}_1,\bm{x}_2,\bm{x}) \in \mathbb{R}^{3n} :
  \bm{x} = \bm{x}_1 + \bm{x}_2,
\label{eq:codebook2:sent}
\\&  \Vert \bm{x}_1 \Vert^2 = n P_1, %\text{\rm(satellite, carries $m_1^\prime := m_{11}$)},
\quad  \Vert \bm{x}_2 \Vert^2 = n P_2, %\text{\rm(cloud center, carries $m_2^\prime := (m_0,m_2,m_{10})$)},
\\&  \langle \bm{x}_1 , \bm{x}_2 \rangle = n\rho\sqrt{P_1 P_2}   %\text{\rm(satellite at a fixed ``angle'' $\rho$)}. 
\Big\}.
\end{align}
\label{eq:codebook2}
\end{subequations}
A transmitted codeword $\bm{x}$ in~\eqref{eq:codebook2} satisfies, because of~\eqref{eq:powparams}, 
\begin{subequations}
\begin{align}
\Vert \bm{x} \Vert^2 
  &= \Vert \bm{x}_1 \Vert^2 + \Vert \bm{x}_2 \Vert^2 + 2 \langle \bm{x}_1 , \bm{x}_2 \rangle
\\&= n P_1 + n P_2 + 2 n\rho\sqrt{P_1 P_2} 
%\\&= n P_1(1-\rho^2)+ n ( \sqrt{P_2} + \rho \sqrt{P_1} )^2 
= n P,
\label{totalpow}
\end{align}
\end{subequations}
i.e., codewords in $\mathcal{D}_n(\rho,P_1,P_2)$ meet the power constraint with equality.
The codewords are chosen independently uniformly at random on their respective power sphere.

\paragraph{Threshold Decoders} 
The channel transition probabilities are $W_j(y|x) = \Ncal\left(y; x, \sigma_j^2 \right),$ $j\in[2]$. 
Let $P_{\bm{X}_2,\bm{X}}$ be the joint distribution induced by the codebook generation, namely\vspace*{-.5cm}
%Define the following distributions induced by the random code construction 
\begin{subequations}
\begin{align}
  & P_{\bm{X}_2,\bm{X}}(\bm{u},\bm{x}) 
  = P_{\bm{X}_2}(\bm{u}) P_{\bm{X}|\bm{X}_2}(\bm{x}|\bm{u})
\label{eq:PX2code}    
\\&= 
\frac{\delta(\Vert \bm{u} \Vert^2 - n P_2)}{S_{n}\big(\sqrt{n P_2}\big)} \
\\&\cdot  \frac{\delta(\Vert \bm{x} - \bm{u} \Vert^2 - n P_1, \ \langle \bm{x} - \bm{u} , \bm{u} \rangle - n\rho\sqrt{P_1 P_2})}{\sqrt{nP} \ S_{n-1}\left( \sqrt{n (1-\rho^2) P_1} \right)},
\label{eq:PX|X2code} 
\end{align}
where the function $S_{n}(\cdot)$ was defined in~\eqref{eq:def n-shpere surface}, which induces 
\begin{align}
P_{\bm{X}}(\bm{x}) = \int P_{\bm{X}_2,\bm{X}}(\bm{u},\bm{x}) \ {\rm d} \bm{u}
 = \frac{\delta(\Vert \bm{x} \Vert^2 - n P)}{S_{n}\big(\sqrt{n P}\big)}.
\label{eq:PXcode} 
\end{align}
Thus for $j\in[2]$ we can compute
\vspace{-0.5em}
\begin{align}
 &P_{\bm{Y}_j|\bm{X}_2}(\bm{y}|\bm{u})
  %= P_{\bm{X}|\bm{X}_2}W_k^n(\bm{y}|\bm{u})
  = \int P_{\bm{X}|\bm{X}_2}(\bm{x}|\bm{u}) W_j^n(\bm{y}|\bm{x}) \ {\rm d} \bm{x}, 
%\\&= \int  \frac{\delta(\Vert \bm{x} - \bm{u} \Vert^2 - n P_1, \ \langle \bm{x}-\bm{u} , \bm{u} \rangle - n\rho\sqrt{P_1 P_2})}{S_{n-1}(\sqrt{n (1-\rho^2) P_1)}} \frac{1}{(2\pi\sigma_k)^{n/2}} \mathrm{e}^{-\Vert \bm{y} - \bm{x} \Vert^2 /(2\sigma_k^2)} \ {\rm d} \bm{x},
%
\vspace{-0.5em}
\\&P_{\bm{Y}_j}(\bm{y})
  %= P_{\bm{X}}W_k^n(\bm{y}) 
  = \int P_{\bm{X}}(\bm{x}) W_j^n(\bm{y}|\bm{x}) \ {\rm d} \bm{x},
%\\&= \int  {\red ???} \ \frac{1}{(2\pi\sigma_k)^{n/2}} \mathrm{e}^{-\Vert \bm{y} - \bm{x} \Vert^2 /(2\sigma_k^2)} \ {\rm d} \bm{x}.
\end{align}
\label{eq:join pdf induced by code}
\vspace{-0.5em}
for $P_{\bm{X}|\bm{X}_2}$ in~\eqref{eq:PX|X2code} and $P_{\bm{X}}$ in~\eqref{eq:PXcode}.
\end{subequations}

\begin{subequations}
In the following we shall use the $n$-fold product of %the following distribution
\begin{align}
&Q_{X_2,X,Y_j}(u,x,y) %X_2,X,Y_
  = Q_{X_2}(u) Q_{X|X_2}(x|u) W_j(y|x) 
\\&= \Ncal\left(
\begin{bmatrix} u \\ x \\ y \\ \end{bmatrix}; \begin{bmatrix} 0 \\ 0 \\ 0 \\ \end{bmatrix}, 
\begin{bmatrix} P_2 & P_2\xi & P_2\xi \\ P_2\xi & P & P \\ P_2\xi & P & P+\sigma_j^2 \\ \end{bmatrix}
%\Sigma_j(\rho) 
\right), 
\label{eq:limitinggaussian joint}
\end{align}
%%%where the covariance matrix (we do not explicitly state the dependance on $(\rho,P_1,P_2)$) is
%%%\begin{align}
%%%\Sigma_j&:= 
%%%\begin{bmatrix} P_2 & P_2\xi & P_2\xi \\ P_2\xi & P & P \\ P_2\xi & P & P+\sigma_j^2 \\ \end{bmatrix},
%%%% \ 
%%%%\begin{matrix}
%%%%j\in[2] \\
%%%%\xi=1+\rho\sqrt{P_1/P_2} \\
%%%%P = (1-\rho^2)P_1+ \xi^2 P_2\\
%%%%\end{matrix},
%%%\end{align}
%%%%which implies 
%%%%\begin{align}
%%%% Q_{k}(x,y|u) %X_2,X,Y_
%%%%  &= Q_{X|X_2}(x|u) W_k(y|x)
%%%%\\&= \Ncal\left(
%%%%\begin{bmatrix} x \\ y \\ \end{bmatrix}; 
%%%%\frac{\sqrt{P_2}+\rho\sqrt{P_1}}{\sqrt{P_2}} u \begin{bmatrix} 1 \\ 1 \\ \end{bmatrix},  
%%%%\begin{bmatrix} (1-\rho^2) P_1 & (1-\rho^2) P_1 \\ (1-\rho^2) P_1 & (1-\rho^2)P_1+\sigma_k^2 \\ \end{bmatrix}\right).
%%%%\end{align}
whose (conditional) marginals are
\begin{align}
   Q_{Y_j|X}(y|x) &= \Ncal\left(y; x, \sigma_j^2 \right) = W_j(y|x),
\label{eq:limitinggaussian Y given X}
\\ Q_{Y_j|X_2}(y|u) &= \Ncal\left(y; \xi u , (1-\rho^2)P_1+\sigma_j^2 \right),
\label{eq:limitinggaussian Y given U}
\\ Q_{Y_j}(y) &= \Ncal\left(y; 0, P+\sigma_j^2 \right).
\label{eq:limitinggaussian Y}
\end{align}
\label{eq:limitinggaussian}
\end{subequations}

%Recall $M_1 = M_{10} M_{11}$, by rate splitting in~\eqref{eq:rate split}.   (from~\eqref{eq:satellite})  (from~\eqref{eq:cloud center})
From the rate split, let $M_1^\prime := M_{11}$and $M_2^\prime := M_{10} M_0 M_2$, therefore $M_0 M_1 M_2 = M_1^\prime M_2^\prime$, and 
\begin{align}
  R_{j,n}^\prime := \frac{1}{n}\ln(M_{j}^\prime), \ j\in[2].
\label{eq:def R{j,n}'}
\end{align}
Also define the mutual information densities 
\begin{align}
  & i_{j,2}\big(\bm{y}; \bm{x}(m_1^\prime,m_2^\prime)\big) := 
  \frac{1}{n}\ln\frac{W_j^n(\bm{y}|\bm{x}(m_1^\prime,m_2^\prime))}{Q^n_{Y_j}(\bm{y})}, 
%\\&\text{ with $Q_{Y_j}$ defined in~\eqref{eq:limitinggaussian Y}}, 
\label{eq:def i_{j,2}}
\\& \hspace{-.3cm}
i_{j,1}\big(\bm{y}; \bm{x}(m_1^\prime,m_2^\prime)|\bm{x}_2(m_2^\prime)\big) := 
  \frac{1}{n}\ln\frac{W_j^n(\bm{y}|\bm{x}(m_1^\prime,m_2^\prime))}{Q^n_{Y_j|X_2}(\bm{y}|\bm{x}_2(m_2^\prime))}, 
%\\&\text{ with $Q_{Y_j|X_2}$ defined in~\eqref{eq:limitinggaussian Y given U}}, 
\label{eq:def i_{j,1}}
\\& i_{j,0}\big(\bm{y}; \bm{x}_2(m_2^\prime)\big) 
  := \frac{1}{n}\ln\frac{Q^n_{Y_j|X_2}(\bm{y}|\bm{x}_2(m_2^\prime))}{Q^n_{Y_j}(\bm{y})}
\label{eq:def i_{j,0}}
\\&\quad=i_{j,2}\big(\bm{y}; \bm{x}(m_1^\prime,m_2^\prime)\big) - i_{j,1}\big(\bm{y}; \bm{x}(m_1^\prime,m_2^\prime)|\bm{x}_2(m_2^\prime)\big),
\notag
\end{align} 
where $Q^n$ denotes the $n$-fold product of the distribution $Q$. %defined in~\eqref{eq:limitinggaussian}.
We employ threshold decoders.
%RX1:
Receiver~1 looks for a unique pair $(m_1^\prime, m_2^\prime)\in [M_1^\prime] \times [M_2^\prime]$ that satisfies
\begin{align}
\begin{cases}
   i_{1,2}\big(\bm{Y}_1; \bm{x}(m_1^\prime,m_2^\prime)\big) 
   &> R_{1,n}^\prime + R_{2,n}^\prime + \gamma                       
   % \frac{1}{n}\ln(M_1^\prime M_2^\prime) + \gamma,
\\ i_{1,1}\big(\bm{Y}_1; \bm{x}(m_1^\prime,m_2^\prime)|\bm{x}_2(m_2^\prime)\big)
   &> R_{1,n}^\prime + \gamma                       
   % \frac{1}{n}\ln(M_1^\prime) + \gamma;
\end{cases};
\label{eq:def decRX1}
\end{align} 
for some $\gamma$; if none or more than one pair of indices are found in~\eqref{eq:def decRX1}, receiver~1 declares an error.
%RX2:
Receiver~2 looks for a unique $m_2^\prime \in [M_2^\prime]$ that satisfies
\begin{align}
  i_{2,0}\big(\bm{Y}_2; \bm{x}_2(m_2^\prime) \big) 
  &> R_{2,n}^\prime + \gamma                       
  % \frac{1}{n}\ln(M_2^\prime) + \gamma;
\label{eq:def decRX2};
\end{align} 
if none or more than one index is found in~\eqref{eq:def decRX2}, receiver~2 declares an error.

\paragraph{Performance Analysis for $\alpha\in(0,1)$} 
%Here we assume that the noises are independent.
The average probability of error, averaged over the messages and over the random code construction, is bounded similarity to the standard typicality decoder~\cite{Cover:2006} as
%. Namely, the probability of error $\epsilon_n$ is super bounded by
\begin{subequations}
\begin{align}
%=\Pr\left[
%  &\left\{ 
%  \frac{1}{n}\ln\frac{W_1^n(\bm{Y}_1|\bm{X})}{Q_{Y_1|x_2}^n(\bm{Y}_1|\bm{X}_2)} > \frac{1}{n}\ln(M_1^\prime) + \gamma,
%  \frac{1}{n}\ln\frac{W_1^n(\bm{Y}_1|\bm{X})}{Q_{Y_1}^n(\bm{Y}_1)} > \frac{1}{n}\ln(M_1^\prime M_2^\prime) + \gamma
%  \right\}^c \cup
%\\&\left\{ 
%  \frac{1}{n}\ln\frac{W_1^n(\bm{Y}_1|\bm{x}(m_1^\prime,m_2^\prime))}{Q_{Y_1|x_2}^n(\bm{Y}_1|\bm{x}_2)(m_2^\prime))} > \frac{1}{n}\ln(M_1^\prime) + \gamma,
%  \frac{1}{n}\ln\frac{W_1^n(\bm{Y}_1|\bm{x}(m_1^\prime,m_2^\prime))}{Q_{Y_1}^n(\bm{Y}_1)} > \frac{1}{n}\ln(M_1^\prime M_2^\prime) + \gamma,
%  \ \text{for some $(m_1^\prime,m_2^\prime)\not=(1,1,1,1)$}
%  \right\} \cup
%\\&\left\{ \frac{1}{n}\ln\frac{Q_{Y_2|x_2}^n(\bm{Y}_2|\bm{X}_2)}{Q_{Y_2}^n(\bm{Y}_1)} > \frac{1}{n}\ln(M_2^\prime) + \gamma 
%  \right\}^c \cup
%\\&\left\{ \frac{1}{n}\ln\frac{Q_{Y_2|x_2}^n(\bm{Y}_2|\bm{x}_2)(m_2^\prime))}{Q_{Y_2}^n(\bm{Y}_1)} > \frac{1}{n}\ln(M_2^\prime) + \gamma
%  \ \text{for some $(m_2^\prime)\not=(1,1,1)$}
%   \right\} 
%\\&\Big| \text{$(m_0,m_2,m_{10},m_{11}) = (1,1,1,1)$ sent}
%\right]
%\\&
&\epsilon_n \leq
1-\Pr
  \left[ 
  \begin{cases}
  i_{2,0}\big(\bm{Y}_2; \bm{X}_2 \big)             > R_{2,n}^\prime                  + \gamma \\
  i_{1,1}\big(\bm{Y}_1; \bm{X}| \bm{X}_2 \big) > R_{1,n}^\prime                  + \gamma \\
  i_{1,2}\big(\bm{Y}_1; \bm{X}\big)                > R_{1,n}^\prime + R_{2,n}^\prime + \gamma \\
  \end{cases} \!\!\!\!
  \right]_{P_0} % = P_{\bm{X}_2,\bm{X}} Q_{Y_1|X}^n Q_{Y_2|X}^n} %P_{\bm{X}_2,\bm{X}} W_1^n W_2^n = 
  \label{eq:epsn:tx not typical}
\\&+K_{2} M_2^\prime \Pr
  \left[
  i_{2,0}\big(\bm{Y}_2; \bm{X}_2 \big) > R_{2,n}^\prime + \gamma
  \right]_{P_2} % = P_{\bm{X}_2,\bm{X}}Q_{Y_2}^n}
  \label{eq:epsn:tx cloud not typical at weak}
\\&
+K_{1} M_1^\prime M_2^\prime \Pr
  \left[
  i_{1,2}\big(\bm{Y}_1; \bm{X}\big) > R_{1,n}^\prime + R_{2,n}^\prime + \gamma
  \right]_{P_1} % = P_{\bm{X}_2,\bm{X}}Q_{Y_1}^n}
  \label{eq:epsn:tx cloud not typical at strong}
\\&+ K_{0} M_1^\prime \Pr
  \left[
  i_{1,1}\big(\bm{Y}_1; \bm{X}| \bm{X}_2 \big) > R_{1,n}^\prime + \gamma
  \right]_{P_3}. % = P_{\bm{X}_2,\bm{X}}Q_{Y_1|X_2}^n}
  \label{eq:epsn:tx satellite not typical}
\end{align}
\label{eq:epsn:tx}
\end{subequations}
Note that there is no ``power constraint violation'' probability in~\eqref{eq:epsn:tx} because we picked the codewords from the set $\mathcal{D}_n(\rho,P_1,P_2)$ in~\eqref{eq:codebook2} to satisfy the power constraint with equality.
In particular we have:

$\bullet$
Eq\eqref{eq:epsn:tx satellite not typical} relates to the event that the receiver~1 has decoded correctly the transmitted cloud center but not the satellite.
The probability is computed from the distribution
\begin{align}
%\Pr_{\bm{X}_2,\bm{X},\bm{Y}_1}(\bm{u},\bm{x},\bm{y}_1) 
P_3 := P_{\bm{X}_2}(\bm{u})P_{\bm{X}|\bm{X}_2}(\bm{x}|\bm{u}) Q_{Y_1|X_2}^n(\bm{y}_1|\bm{u}). % X - U - Y
\end{align}
The factor
\begin{align}
K_{0} = 27~\sqrt{\frac{\pi}{8}} \frac{1+2\gamma_1}{\sqrt{1+4\gamma_1}},
\label{eq:RN K3}
\end{align} 
is the penalty for changing the measure from $P_{\bm{Y}_1|\bm{X}_2}$ to $Q_{Y_1|X_2}^n$, as proven in Lemma~\ref{lemma:K0}. 
Overall, as proven in Lemma~\ref{lemma:han-type bounds} in eq\eqref{eq:RN Y1 given X2 Han's bound}, we have
\begin{align}
\text{\rm eq}\eqref{eq:epsn:tx satellite not typical} 
\leq K_{0} e^{-n \gamma}
\stackrel{\text{for} \ \gamma = \frac{\ln(n)}{2n}}{=} \frac{K_{0}}{\sqrt{n}}.
\end{align}

$\bullet$
Eq\eqref{eq:epsn:tx cloud not typical at strong} relates to the event that receiver~1 has not decoded correctly the transmitted cloud center, and thus also not the satellite. 
The probability is computed from the distribution
\begin{align}
%\Pr_{\bm{X}_2,\bm{X},\bm{Y}_1}(\bm{u},\bm{x},\bm{y}) 
P_1 := P_{\bm{X}_2}(\bm{u})P_{\bm{X}|\bm{X}_2}(\bm{x}|\bm{u}) Q_{Y_1}^n(\bm{y}).
\end{align}
The factor $K_{1}$ is the penalty for changing the measure from $P_{\bm{Y}_1}$ to $Q_{Y_1}^n$, as proven in Lemma~\ref{lemma:Kj} for $j=1$.
Overall, as proven in Lemma~\ref{lemma:han-type bounds} in eq\eqref{eq:RN Y1 Han's bound}, we have
\begin{align}
\text{\rm eq}\eqref{eq:epsn:tx cloud not typical at strong} 
\leq K_{1} e^{-n \gamma}
\stackrel{\text{for} \ \gamma = \frac{\ln(n)}{2n}}{=} \frac{K_{1}}{\sqrt{n}}.
\label{eq:RN K1}
\end{align}
%as proven in Section~\ref{para:han-type bounds}, eq\eqref{eq:RN Y1 Han's bound}.

$\bullet$
Eq\eqref{eq:epsn:tx cloud not typical at weak} relates to the event that receiver~2 has not decoded correctly the transmitted cloud center.
The probability is computed from the distribution
\begin{align}
%\Pr_{\bm{X}_2,\bm{X},\bm{Y}_2}(\bm{u},\bm{x},\bm{y})
P_2 := P_{\bm{X}_2}(\bm{u})P_{\bm{X}|\bm{X}_2}(\bm{x}|\bm{u}) Q_{Y_2}^n(\bm{y}).
\end{align} 
The factor $K_{2}$ is because we changed the measure from $P_{\bm{Y}_2}$ to $Q_{Y_2}^n$, as proven in Lemma~\ref{lemma:Kj} for $j=2$.
Overall, as proven in Lemma~\ref{lemma:han-type bounds} in eq\eqref{eq:RN Y2 Han's bound}, we have
\begin{align}
\text{\rm eq}\eqref{eq:epsn:tx cloud not typical at weak}
\leq K_{2} e^{-n \gamma}
\stackrel{\text{for} \ \gamma = \frac{\ln(n)}{2n}}{=} \frac{K_{2}}{\sqrt{n}}.
\label{eq:RN K2}
\end{align}

$\bullet$
Eq\eqref{eq:epsn:tx not typical} relates to the event that the transmitted codeword does not pass the threshold decoder tests.
The probability is computed from the distribution 
\begin{align}
  P_0
  %&\Pr_{\bm{X}_2,\bm{X},\bm{Y}_1,\bm{Y}_2}(\bm{u},\bm{x},\bm{y}_1,\bm{y}_2)
  %\\
  &:= P_{\bm{X}_2}(\bm{u})P_{\bm{X}|\bm{X}_2}(\bm{x}|\bm{u}) W_1^n(\bm{y}_1|\bm{x}) W_2^n(\bm{y}_2|\bm{x}),
%\\&= P_{\bm{X}_2}(\bm{u})P_{\bm{X}|\bm{X}_2}(\bm{x}|\bm{u}) Q_{Y_1|X}^n(\bm{y}_1|\bm{x}) Q_{Y_2|X}^n(\bm{y}_2|\bm{x}),
\end{align}
since the noises are assumed to be independent. 
\begin{figure*}
\begin{tcolorbox}[colback=white]
\begin{subequations} 
We now upper bound the probability in~\eqref{eq:epsn:tx not typical} for a fixed pair $(\bm{x}_2,\bm{x})$ as
\begin{align}
  &\Pr\left[ 
\begin{bmatrix} 
i_{2,0}\big(\bm{Y}; \bm{x}_2\big)\\
i_{1,1}\big(\bm{Y}; \bm{x}|\bm{x}_2\big)\\ 
i_{1,2}\big(\bm{Y}; \bm{x}\big)\\
\end{bmatrix}
%%%=
%%%\sum_{t\in[n]} 
%%%\begin{bmatrix} \ln \frac{Q_{Y_2|X_2}(Y_{2,t}|x_{2,t})}{Q_{Y_2}(Y_{2,t})} \\ 
%%%                \ln \frac{W_1(Y_{1,t}|x_t)}{Q_{Y_1|X_2}(Y_{1,t}|x_{2,t})} \\
%%%                \ln \frac{W_1(Y_{1,t}|x_t)}{Q_{Y_1}(Y_{1,t})} \\ 
%%%\end{bmatrix}
- \bm{\mu}(\alpha)-\bm{\mu}(\bm{x}_2,\bm{x})
>
\begin{bmatrix} R_{2,n}^\prime + \gamma\\ 
                R_{1,n}^\prime + \gamma\\ 
                R_{1,n}^\prime+R_{2,n}^\prime + \gamma\\
\end{bmatrix} 
- \bm{\mu}(\alpha)-\bm{\mu}(\bm{x}_2,\bm{x})
\right]
\label{eq:before BE}
\\&\geq 
\Pr\left[ 
\bm{Z} %\sim \Ncal\left(\bm{0}_3; \bm{V}(\alpha) + \bm{V}(\bm{x}_2,\bm{x})\right) 
>
\sqrt{n}
\begin{bmatrix} R_{2,n}^\prime + \gamma\\ 
                R_{1,n}^\prime + \gamma\\ 
                R_{1,n}^\prime+R_{2,n}^\prime + \gamma\\
\end{bmatrix} 
-\sqrt{n}\bm{\mu}(\alpha)
-\sqrt{n}\bm{\mu}(\bm{x}_2,\bm{x})
\right]_{\bm{Z}\sim \Ncal\left(\bm{0}_3; \bm{V}(\alpha) + \bm{V}(\bm{x}_2,\bm{x})\right)} 
- \frac{B}{\sqrt{n}}
\label{eq:BE B term}
\\&= 
\Pr\left[ 
\bm{Z} 
<
-\sqrt{n}
\begin{bmatrix} R_{2,n}^\prime + \gamma\\ 
                R_{1,n}^\prime + \gamma\\ 
                R_{1,n}^\prime+R_{2,n}^\prime + \gamma\\
\end{bmatrix} 
+\sqrt{n}\bm{\mu}(\alpha)
+\sqrt{n}\bm{\mu}(\bm{x}_2,\bm{x})
\right]_{\bm{Z}\sim \Ncal\left(\bm{0}_3; \bm{V}(\alpha) + \bm{V}(\bm{x}_2,\bm{x})\right)} 
- \frac{B}{\sqrt{n}}
\\&\stackrel{ \substack{ \gamma = \frac{\ln(n)}{2n} \\ (\bm{x}_1,\bm{x}_2,\bm{x}) \in \mathcal{D}_n(\rho,P_1,P_2) }}{=}
\Psi\left(   -\sqrt{n}
\begin{bmatrix} R_{2,n}^\prime \\ 
                R_{1,n}^\prime \\ 
                R_{1,n}^\prime+R_{2,n}^\prime \\
\end{bmatrix} 
+\sqrt{n}\bm{\mu}(\alpha)
%--%+\sqrt{n}\bm{\mu}(\bm{x}_2,\bm{x})
-\frac{\ln(n)}{2\sqrt{n}}\bm{1}
; \ \bm{V}(\alpha) 
%--%+ \bm{V}(\bm{x}_2,\bm{x})
\right)
- \frac{B}{\sqrt{n}},
\end{align}
where the vectors $\bm{\mu}(\alpha)$ and $\bm{\mu}(\bm{x}_2,\bm{x})$ are defined in~\eqref{eq:mu alpha} and~\eqref{eq:mu x}, respectively;
where the matrices $\bm{V}(\alpha)$ and $\bm{V}(\bm{x}_2,\bm{x})$ are defined in~\eqref{eq:def Valpha} and~\eqref{eq:def Vx}, respectively; 
where in~\eqref{eq:BE B term} we used the multi-variate Berry-Essen theorem, with $B$ a bounded constant that is specified in Lemma~\ref{lemma:BE}; and
% and is bounded because for $\alpha\in(0,1)$ we have $\bm{V}(\alpha) + \bm{V}(\bm{x}_2,\bm{x}) \succ \bm{0}_{3 \times 3}$, that is, the minimum eigenvalue of the dispersion matrix is strictly positive.
where the function $\Psi(\cdot,\cdot)$ was defined in~\eqref{eq:def gaussianCDF}.
We note that any $(\bm{x}_1,\bm{x}_2,\bm{x}) \in \mathcal{D}_n(\rho,P_1,P_2)$ satisfies $\bm{\mu}(\bm{x}_2,\bm{x}) = \bm{0}_3$, and $\bm{V}(\bm{x}_2,\bm{x}) = \bm{0}_{3\times 3}$
%, so the bound in~\eqref{eq:epsn:tx satellite not typical: step1} is valid also when we average with respect to the random codebook generation by setting $\bm{\mu}(\bm{x}_2,\bm{x}) = \bm{0}_3$, and $\bm{V}(\bm{x}_2,\bm{x}) = \bm{0}_{3\times 3}$.
\label{eq:epsn:tx satellite not typical: step1} 
\end{subequations} 
\end{tcolorbox}
\end{figure*}
%==========================================================================================
%==========================================================================================
%==========================================================================================
%
%
Overall, by the multi-dimensional Berry-Essen theorem \cite[Theorem 11]{MACDMS} with $\gamma = {\ln(n)}/{2n}$, we have that the probability on the RHS of~\eqref{eq:epsn:tx not typical} can be upper bounded as proved in~\eqref{eq:epsn:tx satellite not typical: step1} at the top of the next page.
%for $\bm{\mu}(\bm{x}_2,\bm{x}) = \bm{0}_3$ and $\bm{V}(\bm{x}_2,\bm{x}) = \bm{0}_{3\times 3}$. 
In our derivation we used the first and second order moments of the information density vector 
\begin{align}
\bm{i} := 
\begin{bmatrix} 
i_{2,0}\big(\bm{Y}_2; \bm{x}_2(m_2^\prime)\big)\\
i_{1,1}\big(\bm{Y}_1; \bm{x}(m_1^\prime,m_2^\prime)|\bm{x}_2(m_2^\prime)\big)\\ 
i_{1,2}\big(\bm{Y}_1; \bm{x}(m_1^\prime,m_2^\prime)\big)\\
\end{bmatrix},
\label{eq:def idensityvector}
\end{align}
conditioned on a given codeword pair $(\bm{x}(m_1^\prime,m_2^\prime), \bm{x}_2(m_2^\prime))$ chosen from $\mathcal{D}_n(\rho,P_1,P_2)$. 
In~\eqref{eq:def idensityvector} we have sums of independent random variables of the following type, where $Y_{j,t}$ is the channel output at time $t\in[n]$ at receiver $j\in[2]$ 
\begin{subequations} 
\begin{align}
  \ln\frac{W_j(Y_{j,t}|x_t)}{Q_{Y_j|X_2}(Y_{j,t}|u_t)} 
  &= \Csf\left(\alpha \gamma_j\right)
  + \frac{\zeta_{j,t}^2 - N_{j,t}^2\ \alpha \gamma_j}{2(1+\alpha \gamma_j)}
  + \left.\frac{\zeta_{j,t} N_{j,t}}{1+\alpha \gamma_j}\right., %|_{N_j = \frac{Y_j-x}{\sigma_j} \sim \Ncal(0,1), \zeta_j = \frac{x-\xi u  }{\sigma_j}},
\\ 
  \ln\frac{W_j(Y_{j,t}|x_t)}{Q_{Y_j}(Y_{j,t})} 
  &= \Csf\left(\gamma_j\right) 
  + \frac{\nu_{j,t}^2 - N_{j,t}^2\  \gamma_j}{2(1+ \gamma_j)} 
  + \left.\frac{\nu_{j,t} N_{j,t}}{1+ \gamma_j} \right., %|_{N_j = \frac{Y_j-x}{\sigma_j} \sim \Ncal(0,1), \nu_j = \frac{x}{\sigma_j}},
\end{align}
where we introduced the normalized quantities
\begin{align}
  & N_{j,t}     := \frac{Y_{j,t}-x_t}{\sigma_j} \sim \Ncal(0,1), %noise
\\& \zeta_{j,t} := \frac{x_t-\xi u_t}{\sigma_j}, %satellite
%\\& 
\quad
\nu_{j,t}   := \frac{x_t}{\sigma_j}.   %whole codeword
\end{align} 
\label{eq:i's for fixed codeword}
\end{subequations}
By summing over $t\in[n]$ in~\eqref{eq:i's for fixed codeword} and with the shorthand notation $(\bm{x},\bm{x}_2)$ for $(\bm{x}(m_1^\prime,m_2^\prime)\big),\bm{x}_2(m_2^\prime))$, 
we obtain that the means of the random variables in~\eqref{eq:def idensityvector} are 
\begin{subequations} 
\begin{align}
\mathbb{E}  &\left[i_{j,2}\big(\bm{x}+\bm{Z}_j; \bm{x}\big) \right]%_{\bm{Y} = \bm{x}+\bm{Z}_j}
%\\&=\mathbb{E}  \left[ \ln\frac{W_j^n(\bm{Y}_j|\bm{x})}{Q_{Y_j}^n(\bm{Y}_j)} \right] 
  = \Csf\left(\gamma_j\right) 
  + \frac{\Vert \bm{x} \Vert^2 / \sigma_j^2 - n \gamma_j}{2n(1+ \gamma_j)},
\label{eq:i's for fixed codeword mean 0}
\\
\mathbb{E}  &\left[i_{j,1}\big(\bm{x}+\bm{Z}_j; \bm{x}|\bm{x}_2\big) \right]%_{\bm{Y} = \bm{x}+\bm{Z}_j}
%\\&=\mathbb{E} \left[ \ln\frac{W_j^n(\bm{Y}_j|\bm{x})}{Q_{Y_j|X_2}^n(\bm{Y}_j|\bm{x}_2)} \right] 
\\&= \Csf\left(\alpha \gamma_j\right) 
   + \frac{\Vert \bm{x}-\xi \bm{x}_2 \Vert^2 / \sigma_j^2  - n\alpha \gamma_j}{2n(1+\alpha \gamma_j)},
\label{eq:i's for fixed codeword mean 1}
\\
\mathbb{E}&\left[i_{j,0}\big(\bm{x}+\bm{Z}_j; \bm{x}_2\big) \right]%_{\bm{Y} = \bm{x}+\bm{Z}_j}
%\\&= \mathbb{E}\left[i_{j,2}\big(\bm{Y}; \bm{x}(m_1^\prime,m_2^\prime)|\bm{x}_2(m_2^\prime)\big) \right]
%\\&\quad- \mathbb{E}\left[i_{j,1}\big(\bm{Y}; \bm{x}(m_1^\prime,m_2^\prime)|\bm{x}_2(m_2^\prime)\big) \right]
   ={\rm eq}\eqref{eq:i's for fixed codeword mean 0} 
  - {\rm eq}\eqref{eq:i's for fixed codeword mean 1},
\label{eq:i's for fixed codeword mean 2}
\end{align}
\label{eq:i's for fixed codeword mean}
\end{subequations} 
and the (co)variances are
\begin{subequations} 
\begin{align}
n\mathrm{Var}&\left[i_{j,2}\big(\bm{x}+\bm{Z}_j; \bm{x}\big) \right]
\\&= \frac{1}{2}\left( \frac{\gamma_j}{1+\gamma_j} \right)^2
   + \frac{\Vert \bm{x} \Vert^2 / \sigma_j^2}{n(1+\gamma_j)^2},
\label{eq:i's for fixed codeword var 0}
\\
n\mathrm{Var}&\left[i_{j,1}\big(\bm{x}+\bm{Z}_j; \bm{x}|\bm{x}_2\big) \right]
\\&= \frac{1}{2}\left( \frac{\alpha \gamma_j}{1+\alpha \gamma_j} \right)^2
   + \frac{\Vert \bm{x}-\xi \bm{x}_2 \Vert^2 / \sigma_j^2}{n(1+\alpha \gamma_j)^2},
\label{eq:i's for fixed codeword var 1}
\\
n\mathrm{Cov}&\left[ i_{j,2}\big(\bm{x}+\bm{Z}_j; \bm{x}\big), \ %\right.\\&\left. \quad \notag 
                     i_{j,1}\big(\bm{x}+\bm{Z}_j; \bm{x}|\bm{x}_2\big) \right]
\\&= \frac{1}{2} \frac{\alpha \gamma_j}{1+\alpha \gamma_j} \frac{\gamma_j}{1+\gamma_j}
  + \frac{\langle \bm{x}- \xi\bm{x}_2 , \bm{x} \rangle / \sigma_j^2}{n(1+\alpha \gamma_j)(1+\gamma_j)},
\label{eq:i's for fixed codeword var covar}
\\
n\mathrm{Var}&\left[ i_{j,0}\big(\bm{x}+\bm{Z}_j; \bm{x}_2\big) \right]%_{\bm{Y} = \bm{x}+\bm{Z}_j}
  = {\rm eq}\eqref{eq:i's for fixed codeword var 0}
   +{\rm eq}\eqref{eq:i's for fixed codeword var 1}
   -2 \cdot {\rm eq}\eqref{eq:i's for fixed codeword var covar},
\label{eq:i's for fixed codeword var 2}
\\
n\mathrm{Cov}&[ i_{1,\ell_1}(\cdots), i_{2,\ell_2}(\cdots)]=0, \forall (\ell_1,\ell_2)\in[0:2]^3,
\label{eq:i's for fixed codeword var crosscovar}
\end{align}
where~\eqref{eq:i's for fixed codeword var crosscovar} follows because the noises at the two receivers are assumed to be independent.
\label{eq:i's for fixed codeword var}
\end{subequations} 
Thus, the information density vector in~\eqref{eq:def idensityvector} has mean $\mathbb{E}[\bm{i} ] = \bm{\mu}(\alpha) +\bm{\mu}(\bm{x}_2,\bm{x})$ with
\begin{subequations}
\begin{align}  
\bm{\mu}(\alpha) &:=
\begin{bmatrix} \Csf\left(\gamma_2\right) 
              - \Csf\left(\alpha \gamma_2\right)  \\ 
                \Csf\left(\alpha \gamma_1\right)  \\
                \Csf\left(\gamma_1\right)  \\ 
\end{bmatrix},
\label{eq:mu alpha}
%\end{align}
\\%and
%\begin{align}
\bm{\mu}(\bm{x}_2,\bm{x}) &:=
\begin{bmatrix} \frac{\Vert \bm{x} \Vert^2 / \sigma_2^2 - n \gamma_2}{n2(1+ \gamma_2)} 
              - \frac{\Vert \bm{x}-\xi \bm{x}_2 \Vert^2 / \sigma_2^2  - n\alpha \gamma_2}{n2(1+\alpha \gamma_2)} \\ 
                \frac{\Vert \bm{x}-\xi \bm{x}_2 \Vert^2 / \sigma_1^2  - n\alpha \gamma_1}{n2(1+\alpha \gamma_1)} \\
                \frac{\Vert \bm{x} \Vert^2 / \sigma_1^2 - n \gamma_1}{n2(1+ \gamma_1)} \\ 
\end{bmatrix},
\label{eq:mu x}
\end{align}
\label{eq:i's for fixed codeword mean vector}
\end{subequations}
and covariance matrix $n\mathrm{Cov}[\bm{i} ]=\bm{V}(\alpha) + \bm{V}(\bm{x}_2,\bm{x})$ with
\begin{subequations}
\begin{align}
\bm{V}(\alpha) &=
\begin{bmatrix}
\bm{V}_2(\alpha) & 0 \\
0 & \bm{V}_1(\alpha) \\
\end{bmatrix},
\label{eq:def Valpha}
\\
\bm{V}_2(\alpha) &:= 
\begin{bmatrix}
\Vsf^\prime(\alpha \gamma_2,\gamma_2)
\end{bmatrix},
\label{eq:def V2alpha}
\\
\bm{V}_1(\alpha) &:=
\begin{bmatrix}
\Vsf(\alpha \gamma_1,\alpha \gamma_1) & \Vsf(\alpha \gamma_1,\gamma_1) \\
\Vsf(\alpha \gamma_1,\gamma_1)        & \Vsf(\gamma_1,\gamma_1)  \\
\end{bmatrix},
\label{eq:def V1alpha}
\end{align}
for $\Vsf^\prime(\cdot,\cdot)$ and $\Vsf(\cdot,\cdot)$ defined in~\eqref{eq:def V11} and~\eqref{eq:def Vxy}, respectively, and
\begin{align}
&\bm{V}(\bm{x}_2,\bm{x}) :=
\begin{bmatrix}
\upsilon_{2,11}+\upsilon_{2,22}-2\upsilon_{2,12} 
  & \!\!\!\! 0 & 0 \\
0 & \!\!\!\! \upsilon_{1,11}
  & \upsilon_{1,12} \\
0 & \!\!\!\! \upsilon_{1,12}
  & \upsilon_{1,22}  \\
\end{bmatrix},
\label{eq:def Vxij}
\\& \upsilon_{j,11} := \frac{\Vert \bm{x}-\xi \bm{x}_2 \Vert^2 / \sigma_j^2}{n(1+\alpha \gamma_j)^2}  - \frac{\alpha \gamma_j}{(1+\alpha \gamma_j)^2},
\\& \upsilon_{j,22} := \frac{\Vert \bm{x} \Vert^2 / \sigma_j^2}{n(1+\gamma_j)^2} - \frac{\gamma_j}{(1+\gamma_j)^2},
\\& \upsilon_{j,12} := \frac{\langle \bm{x}- \xi\bm{x}_2 , \bm{x} \rangle / \sigma_j^2}{n(1+\alpha \gamma_j)(1+\gamma_j)} - \frac{\alpha \gamma_j}{(1+\alpha \gamma_j)(1+\gamma_j)}.
\end{align}
By construction, every codeword satisfies $\bm{\mu}(\bm{x}_2,\bm{x}) = \bm{0}_3$, and $\bm{V}(\bm{x}_2,\bm{x}) = \bm{0}_{3\times 3}$.
\label{eq:i's for fixed codeword var matrix}
\end{subequations}

%==========================================================================================
%==========================================================================================
%==========================================================================================
\begin{figure*}
\begin{tcolorbox}[colback=white]
\begin{subequations}
The probability of error, averaged over the random code construction, can be bounded as
\begin{align}
\epsilon_n
\leq 1 - 
\Psi\left( -\sqrt{n}\begin{bmatrix} R_{2,n}^\prime - \Csf\left(\gamma_2\right) + \Csf\left(\alpha \gamma_2\right) 
                                    \\ R_{1,n}^\prime - \Csf\left(\alpha \gamma_1\right) 
                                    \\ R_{1,n}^\prime+R_{2,n}^\prime - \Csf\left(\gamma_1\right) 
                                    \\ \end{bmatrix} -\frac{\ln(n)}{2\sqrt{n}}\bm{1}; \bm{V}(\alpha) \right)
+ \frac{B+K_{1}+K_{2}+K_{0}}{\sqrt{n}}
\stackrel{ \substack{ \text{to meet constraint} \\
           \text{in Definition~\ref{def: Second Order Region}}
         }}{\leq} \varepsilon,
\label{eq:alomostdone 1}
\\ 
\Longleftrightarrow
\begin{bmatrix} R_{2,n}^\prime - \Csf\left(\gamma_2\right) + \Csf\left(\alpha \gamma_2\right) 
             \\ R_{1,n}^\prime - \Csf\left(\alpha \gamma_1\right) 
             \\ R_{1,n}^\prime+R_{2,n}^\prime - \Csf\left(\gamma_1\right) 
             \end{bmatrix}
 +\frac{\ln(n)}{2n}\bm{1}
\in
 \frac{1}{\sqrt{n}}\Qsf_\text{\rm inv}\left( \varepsilon - \frac{B+K_{1}+K_{2}+K_{0}}{\sqrt{n}}; \bm{V}(\alpha) \right)
\label{eq:alomostdone 2}
\\
\Longleftrightarrow
\begin{bmatrix} R_{2,n}^\prime  
             \\ R_{1,n}^\prime  
             \\ R_{1,n}^\prime+R_{2,n}^\prime 
             \end{bmatrix}
\in
\begin{bmatrix} \Csf\left(\gamma_2\right) - \Csf\left(\alpha \gamma_2\right) 
             \\ \Csf\left(\alpha \gamma_1\right) 
             \\ \Csf\left(\gamma_1\right) 
             \end{bmatrix}
+ \frac{1}{\sqrt{n}}\Qsf_\text{\rm inv}\left( \varepsilon; \bm{V}(\alpha) \right) + O\left(\frac{\ln(n)}{n}\right),
\label{eq:alomostdone 3}
\end{align}
where for~\eqref{eq:alomostdone 1} the function $\Psi(\cdot,\cdot)$ was defined in~\eqref{eq:def gaussianCDF},
      for~\eqref{eq:alomostdone 2} the function $\Qsf_\text{\rm inv}(\cdot;\cdot)$ was defined in~\eqref{eq:def Qinv}, 
      the covariance matrix $\bm{V}(\alpha)$ was defined in~\eqref{eq:def Valpha}, and
where~\eqref{eq:alomostdone 3} follows from the continuity of $\Qsf_\text{\rm inv}(\varepsilon;\cdot)$ in $\varepsilon$ proved similarly to~\cite[Lemma 5 proved in Appendix C]{MACDMS}.
\label{eq:alomostdone}
\end{subequations}
\end{tcolorbox}
\end{figure*}
%==========================================================================================
%==========================================================================================
%==========================================================================================
%
%
$\bullet$
By combining everything together, %and since all codewords in $\mathcal{D}_n(\rho,P_1,P_2)$  satisfy $\bm{\mu}(\bm{x}_2,\bm{x}) = \bm{0}_3$ and $\bm{V}(\bm{x}_2,\bm{x}) = \bm{0}_{3\times 3}$, 
we obtain the relationship in~\eqref{eq:alomostdone} at the top of this page.
\begin{subequations}
The set $\Qsf_\text{\rm inv}\left( \varepsilon;\bm{V}(\alpha) \right)$ in~\eqref{eq:alomostdone 3} for the block diagonal covariance matrix $\bm{V}(\alpha)$ in~\eqref{eq:def Valpha} can be written as
\begin{align}
&\Qsf_\text{\rm inv}\left( \varepsilon; \bm{V}(\alpha) \right)%_{\text{for $\bm{V}(\alpha)$ given in~\eqref{eq:def Valpha}}}
= \big\{ \bm{a}\in \mathbb{R}^3 :\Pr\left[\bm{Z} \leq -\bm{a} \right]
%_{\bm{Z} \sim \Ncal\left( \bm{0}, \bm{V}(\alpha) \right)} 
  \geq 1-\varepsilon \big\}
\\&= \Big\{ \bm{a}\in \mathbb{R}^3 :
a_i = - \sqrt{ [\bm{V}(\alpha)]_{ii} } \, \Qsf^{-1}(\epsilon_i), \ i\in[3],
%\begin{matrix}
%a_1 = - \sqrt{ \Vsf^\prime(\alpha \gamma_2,\gamma_2) } \Qsf^{-1}(\epsilon_2)\\
%a_2 = - \sqrt{ \Vsf(\alpha \gamma_1) } \Qsf^{-1}(\epsilon_{10})\\
%a_3 = - \sqrt{ \Vsf(\gamma_1) } \Qsf^{-1}(\epsilon_{11})\\
%\end{matrix}
\\&
\quad \text{for $(\epsilon_{10},\epsilon_{11},\epsilon_2)\in[0,1]^3$ that satisfy} 
\\&
1-\varepsilon \leq \Pr\left[ G_1 \leq \Qsf^{-1}(\epsilon_2) \right]_{ G_1 \sim \Ncal(0,1) }
\\&\cdot
\Pr\left[ G_2 \leq \Qsf^{-1}(\epsilon_{10}),
          G_3 \leq \Qsf^{-1}(\epsilon_{11}) 
          \right]_{\tiny  \begin{bmatrix} G_2 \\ G_3 \\ \end{bmatrix} \sim \Ncal\left( \bm{0}, \begin{bmatrix} 1 \ r \\ r \ 1 \\ \end{bmatrix}  \right)} 
\\&=
(1-\epsilon_2)\mathsf{F}(\epsilon_{10},\epsilon_{11};r)
%\\&
%\mathsf{F}(\epsilon_{10},\epsilon_{11};r)=1-\epsilon_1,
\Big\},
\end{align}
where $\mathsf{F}(\epsilon_{10},\epsilon_{11};r)$ was defined in~\eqref{eq:sup Fdef}.
\end{subequations}
This proves the achievability of $\Rcal^{\text{\rm(SUP)}}(n,\varepsilon)$ in~\eqref{eq:normal approximation sup} for $\alpha\in(0,1)$.

\paragraph{Performance Analysis for $\alpha=0$}  
%In the case $\alpha=0$, 
Here the step in the above derivation where we used the multivariate Berry-Essen theorem does not hold because the $3 \times 3$ covariance matrix $\bm{V}(0)$ (from $\bm{V}(\alpha)$ in~\eqref{eq:def Valpha} evaluated for $\alpha=0$) has rank 2. In this case, our scheme reduces to a standard point-to-point codebook on the power sphere, that is $\Vert \bm{x}(m_0,m_1,m_2) \Vert^2 = n P$ for all $(m_0,m_1,m_2)\in[M_0]\times[M_1]\times[M_2]$, and where each receiver $j\in[2]$ looks for the triplet $(m_0,m_1,m_2)$ that satisfies $i_{j,2}\big(\bm{y}_j; \bm{x}(m_0,m_1,m_2)\big) > R_{1,n}^\prime + R_{2,n}^\prime + \gamma$. The analysis proceeds as done for $\alpha\in(0,1)$, except that the information density vector has dimension 2 rather than 3.
The resulting region is as in~\eqref{eq:normal approximation sup} for the choice $\beta=1, \alpha=0, (1-\epsilon_{11})(1-\epsilon_2)=1-\varepsilon$ (here $\epsilon_{10}$ does not matter).
%--see also Remark~\ref{rem:CCP}.

\paragraph{Performance Analysis for $\alpha=1$} 
%As for the case $\alpha=0$, 
Here too the $3 \times 3$ covariance matrix $\bm{V}(1)$ (from $\bm{V}(\alpha)$ in~\eqref{eq:def Valpha} evaluated for $\alpha=1$) has rank 2. In this case,
%no message is sent to the user with the lowest SNR, that is,
$R_0=R_2=0$. Our scheme reduces to a standard point-to-point codebook on the power sphere, that is, $\Vert \bm{x}(m_1) \Vert^2 = n P$ for all $(m_1)\in[M_1]$, and where receiver~1 looks for an index $m_1$ that satisfies $i_{1,2}\big(\bm{y}_1; \bm{x}(m_1)\big) > R_{1,n}^\prime + \gamma$. Receiver~2 does not do anything. The analysis proceeds as in the point-to-point case. The resulting region is as in~\eqref{eq:normal approximation sup} for the choice $\beta=0, \alpha=1, \epsilon_{10}=\varepsilon$ (here $\epsilon_{11}$ and $\epsilon_2$ do not matter).

\subsection{Extension to \texorpdfstring{$K$}{K} users}\label{sec:inK}
For simplicity, we only consider private rates and no splitting here.
WLOG we assume $\gamma_1 \geq \gamma_2 \geq \ldots \gamma_K > 0$.

\paragraph{Capacity Region}
The capacity region $\Ccal$ of the $K$-user degraded BC $X \to Y_1 \to Y_2 \ldots \to Y_{K-1} \to Y_K$
is attained by superposition coding, where the $K$ levels of superposition satisfy the Markov chain %(here we can choose the joint distribution of the noise as we see fit as long as the marginals are preserved)
\begin{align}
  & U_K \to U_{K-1} \to \ldots \to U_1 \to X. %\to Y_j, j\in[K]. %1 \to Y_2 \ldots \to Y_{K-1} \to Y_K;
%\\& Y_j = X + Z_j, \  Z_j \sim \Ncal\left(0, \sigma_j^2 \right), \  \mathbb{E}[X^2]\leq P;
%\\& 0< \sigma_1 \leq \sigma_2 \leq \ldots \sigma_{K-1} \leq \sigma_K;
\label{eq:supKusersMC}
\end{align} 
For the AWGN BC, we have
\begin{subequations}
\begin{align}
\Ccal 
%  &= \bigcup_{P_{X,U_1,...,U_K}}  
%\Big\{ (R_1,R_2,\ldots,R_K)\in \mathbb{R}^K_+ : \forall k\in[K]
%\\&
%R_k \leq I(Y_k; U_k|U_{k+1}, \ldots, U_K)
%    %= I(Y_k; X|U_{k+1}, \ldots, U_K) - I(Y_k; X|U_k, U_{k+1}, \ldots, U_K),  
%\Big\}
%\\
&= 
%\hspace*{-.5cm}
\bigcup%_{ \substack{(\alpha_1,\ldots\alpha_K)\in[0,1]^K : \\ \sum_{k\in[K]} \alpha_k = 1}} %\cap_{\forall k\in[K]}
%\hspace*{-.5cm}
\Big\{(R_1,R_2,\ldots,R_K)\in \mathbb{R}^K_+ :  \forall k\in[K]
\label{eq:alphaKuser union}
\\&
   R_k \leq \Csf\Big(\gamma_k\sum_{\ell\in[k]}\alpha_\ell\Big) - \Csf\Big(\gamma_k\sum_{\ell\in[k-1]}\alpha_\ell\Big)   
\Big\},
\end{align} 
where the union in~\eqref{eq:alphaKuser union} is over the ``power splits''
\begin{align}
(\alpha_1,\ldots\alpha_K)\in[0,1]^K : \sum_{\ell\in[K]} \alpha_\ell = 1.
\label{eq:alphaKuser}
\end{align}
The capacity region in~\eqref{eq:alphaKcapacity} is attained, for example, by mutually independent $U_k \sim \Ncal\left(0, \alpha_k P \right), \forall k\in[K],$ and $X = \sum_{k\in[K]}U_k$ in~\eqref{eq:supKusersMC} such that~\eqref{eq:alphaKuser} holds. 
\label{eq:alphaKcapacity}
\end{subequations}

\paragraph{First-Order Superposition Coding Region}
Consider a fixed $(\alpha_1,\ldots\alpha_K)$ as in~\eqref{eq:alphaKuser}.
In order not to clutter the notation next we omit to explicitly state the dependence on $(\alpha_1,\ldots\alpha_K)$ of various quantities, unless necessary or not clear from the context.
%$P_{X,U_1,...,U_K}$ satisfying the Markov chain in~\eqref{eq:supKusersMC}.
For the purpose of developing a second order region, we write the capacity achieving superposition coding region with Gaussian input, where user $j\in[K]$ jointly decodes all the messages intended for the users indexed by $\{j, j+1, \ldots, K\}$, as follows
\begin{align}
%  & U_K^n(m_K) \to U_{K-1}^n(m_K,m_{K-1}) \to \ldots \to U_2^n(m_K,m_{K-1},\ldots,m_2) \to U_1^n(m_K,m_{K-1},\ldots,m_2,m_1) \to X \to Y_j^n,
%\\& (U_K^n(m_K), \ldots, U_j^n(m_K,m_{K-1},\ldots,m_j),Y_j^n) \in \text{JointlyTypicalSet}, \  \forall j\in[K];
%\\&
%\Rcal = \bigcup_{P_{X,U_1,...,U_K}} 
\cap_{j\in[K]}
\begin{cases}
  R_K+R_{K-1} + \ldots + R_j &\leq %I(U_K, U_{K-1},\ldots, U_{j+1},U_j; Y_j) = 
  I_{j,K} -I_{j,j-1}\\
      R_{K-1} + \ldots + R_j &\leq %I(     U_{K-1},\ldots, U_{j+1},U_j; Y_j|U_K) = 
      I_{j,K-1} -I_{j,j-1} \\
 \vdots \\
                R_{j-1}+ R_j &\leq %I(U_{j+1},U_j; Y_j|U_K,\ldots, U_{j+2}) =
                I_{j,j+1} -I_{j,j-1}\\
                         R_j &\leq %I(        U_j; Y_j|U_K,\ldots, U_{j+1}) =
                         I_{j,j} -I_{j,j-1}\\
\end{cases}                         
\label{eq:KBCAWGNSup}
\end{align} 
where $I_{j,\ell}$ is the mutual information at receiver $j\in[K]$ to decodes the messages indexed by $\{1, \ldots, \ell\} : \ell\in[0:K]$ after having removed the effect of the messages indexed by $\{\ell+1, \ldots, K\}$, that is,
\begin{subequations}
\begin{align}
I_{j,\ell} := I(X; Y_j |U_{\ell+1}^K)= \Csf\Big(\gamma_j \sum_{k\in[\ell]} \alpha_k\Big),
\end{align} 
with the convention that $I_{j,0}=0$ and $U_{K+1}^K=\emptyset$, 
which satisfy
\begin{align}
&0= I_{j,0} \leq I_{j,1} \leq I_{j,2} \ldots \leq I_{j,K} =\Csf(\gamma_j). %= I(X; Y_j)
\end{align} 
\end{subequations}
We next we aim to find a second order region for~\eqref{eq:KBCAWGNSup}.

\paragraph{Random Codebook Generation}
For $(\alpha_1,\ldots\alpha_K)$ as in~\eqref{eq:alphaKuser}, define
\begin{subequations} 
\begin{align}
\mathcal{D}_n&(\alpha_1,\ldots,\alpha_K) := \Big\{ 
  (\bm{x}_1,\ldots,\bm{x}_K,\bm{x}) \in \mathbb{R}^{(K+1)n} : 
\\& \bm{x} = \sum_{k\in[K]}\bm{x}_k, \
\label{eq:TXKusers}
\\& 
\langle \bm{x}_{j} , \bm{x}_{\ell}   \rangle = \delta(\ell-j) \, n\alpha_j P, \  \forall (j,\ell)\in[K]^2
\Big\}.
\end{align} 
As in Footnote~\ref{foot:costruction2} for the two-user case, we choose the sub-codeword $\bm{x}_k$ independently and uniformly at random on the power sphere $\mathcal{S}_{n-K+k}(\sqrt{n\alpha_k P})$ and mutually orthogonal.
%to the other sub-codewords. %, i.e., from $\mathcal{D}_n(\alpha_1,\ldots,\alpha_K)$. 
The resulting transmitted codeword in~\eqref{eq:TXKusers} 
%$\bm{x} = \sum_{k\in[K]}\bm{x}_k$ 
meets the power constraint $nP$ with equality. %(see also Appendix~\ref{app:mimic any given Sigma}). 
This construction aims to mimic a choice of independent Gaussian $U_1,\ldots,U_K$ in~\eqref{eq:supKusersMC}.
\label{eq:codebookK}
\end{subequations} 

\paragraph{Threshold Decoding}
Define auxiliary distributions
\begin{align}
Q_{j,\ell}(y | v_{\ell+1}, \ldots, v_{K}) &= \Ncal\Big(y; \sum_{i\in[\ell+1:K]} \!\!\! v_i, \ \sigma_j^2 + P\sum_{i\in[\ell]}\alpha_i \Big), 
\end{align}
for $(j,\ell)\in[K]\times[0:K]$, with the convention $\sum_{i\in[0]}\alpha_i=0$ and $\sum_{i\in[K+1:K]} v_{i} = 0$; with this we have
\begin{align}
Q_{j,0}(y|v_1, \ldots, v_K) = W_j\Big(y \big| \sum_{i\in[K]} v_i\Big). %= \Ncal\left(y; x,  \sigma_j^2 \right);
\end{align}
Let $\gamma = {\ln(n)}/{2n}$.
Receiver $j\in[K]$, upon receiving $\bm{y}_j$, looks for a unique $(m_{j},m_{j+1}\ldots,m_{K}) \in[M_{j}]\times[M_{j+1}]\times\ldots\times[M_{K}]$ such that 
%\begin{itemize}
%\item User $K$: for received $\bm{y}$ at receiver $K$, find unique $m_{K}$ such that
\begin{align}
R_{j}+\ldots+R_{\ell}
&> i_{j,\ell}(\bm{y}_j) - i_{j,j-1}(\bm{y}_j) + \gamma, \ \forall \ell \in[j: K],
\end{align}
where--omitting message indices for readability--we defined
\begin{align}
i_{j,\ell}(\bm{Y}_j)
  &:= \left. \frac{1}{n}\ln \frac{W_j^n(\bm{Y}_j | \bm{x}^n)}{Q_{j,\ell}^n(\bm{Y}_j | \bm{x}_{\ell+1}^n, \ldots, \bm{x}_{K}^n)}\right|_{\bm{Y}_j=\bm{x}+\bm{Z}_j}.
\label{eq:def i_{j,ell}}
\end{align}
%by which we mean that we condition on a transmitted codeword and average with respect to the Gaussian noise (i.e., given the conditioning on the transmitted codeword $\bm{x}$, we have $\bm{Y}=\bm{x}+\bm{Z}$).

\paragraph{Performance Analysis}
\begin{subequations}
Define
%For later use, define
%\begin{align}
%  &\bm{s}_{j,\ell} :=\frac{\bm{x}-\sum_{i\in[\ell+1:K]} \bm{x}_{i}}{\sigma_j} = \frac{\sum_{i\in[\ell]} \bm{x}_{i}}{\sigma_j},
%\end{align}
%with power %{\red !! CHECK IT!!} and the inner products are zero {\red !! CHECK IT!!}, where we 
\begin{align}
  &P_{j,\ell} := %\frac{1}{n}\Vert \bm{s}_{j,\ell} \Vert^2 =  
  \gamma_j \sum_{i\in[\ell]}\alpha_i, 
\end{align}
which satisfy
\begin{align}
0= P_{j,0} \leq P_{j,1} \leq \ldots \leq P_{j,K} = \gamma_j. %\equiv \frac{1}{\sigma_j^2}=\text{``SNR on channel $j$''}.
\end{align}
The analysis proceeds as for the two-user case but with information density vectors of larger dimension.
For receiver $j\in[K]$, consider the $(K-j+2)$-dimensional information density random vector 
\begin{align}
[i_{j,j-1}(\bm{Y}_j); i_{j,j}(\bm{Y}_j); \ldots; i_{j,K}(\bm{Y}_j)],
\end{align}
whose mean vector and covariance matrix conditioned on a transmitted codeword from $\mathcal{D}_n(\alpha_1,\ldots,\alpha_K)$--omitting to explicitly state the conditioning for readability--have entries
%of the information densities in~\eqref{eq:def i_{j,ell}}  
\begin{align}
\mathbb{E}[i_{j,\ell}(\bm{Y}_j)]
%  &= \ln\left(1+ P_{j,\ell} \right)
%  + 0;
%\\
&=\Csf\left(P_{j,\ell} \right);
\\
n \mathrm{Var}[i_{j,\ell}(\bm{Y}_j)] 
%  &= \frac{1}{2}\left(\frac{P_{j,\ell}}{1+P_{j,\ell}} \right)^2
%   + \frac{P_{j,\ell} }{(1+P_{j,\ell})^2}
%   = \Vsf \left(P_{j,\ell},P_{j,\ell}\right)
%\\
&=\Vsf \left(P_{j,\ell}\right);
\\
n \mathrm{Cov}[i_{j,\ell_1}(\bm{Y}_j), i_{j,\ell_2}(\bm{Y}_j)]
%  &= \frac{1}{2}\left(\frac{P_{j,\ell_1}}{1+P_{j,\ell_1}} \right) \left(\frac{P_{j,\ell_2}}{1+P_{j,\ell_2}} \right)
%   + \frac{P_{j,\min\{\ell_1, \ell_2\}}}{(1+P_{j,\ell_1})(1+P_{j,\ell_2})}
%\\
&=\Vsf \left(P_{j,\min\{\ell_1, \ell_2\}},P_{j,\max\{\ell_1, \ell_2\}}\right),
\\
n \mathrm{Cov}[i_{a,\ell_1}(\bm{Y}_a), i_{b,\ell_2}(\bm{Y}_b)] &=0, \ a\not=b.
\end{align}
\label{eq:mu and Sigma of i(j,ell)}
\end{subequations}
\begin{subequations}
Next, for receiver $j\in[K]$, from the means and covariances in~\eqref{eq:mu and Sigma of i(j,ell)}, we evaluate the mean vector
\begin{align}
\mathbb{E}[\bm{i}_j] = \bm{\mu}_j(\alpha_1,\ldots\alpha_K),
\end{align}
and the covariance matrix
\begin{align}
n \mathrm{Cov}[\bm{i}_j] = \bm{V}_j(\alpha_1,\ldots\alpha_K),
\end{align} 
of the $(K-j+1)$-dimensional random vector
\begin{align}
\bm{i}_j := [i_{j,j}(\bm{Y}_j) - i_{j,j-1}(\bm{Y}_j); \ldots; i_{j,K}(\bm{Y}_j) - i_{j,j-1}(\bm{Y}_j)], 
\end{align}
whose entries satisfy 
\begin{align}
  &\mathbb{E}[i_{j,\ell}(\bm{Y}_j) -i_{j,j-1}(\bm{Y}_j)] = \Csf\left(P_{j,\ell} \right)-\Csf\left(P_{j,j-1} \right),
\end{align} 
and for $j-1 < \min(\ell_1,\ell_2)$
\begin{align}
  &n \mathrm{Cov}[i_{j,\ell_1}(\bm{Y}_j) -i_{j,j-1}(\bm{Y}_j), \ i_{j,\ell_2}(\bm{Y}_j) - i_{j,j-1}(\bm{Y}_j)]
%\\&=
% \mathrm{Cov}[i_{j,\ell_1} , i_{j,\ell_2}] 
%+\mathrm{Var}[i_{j,j-1}]
%-\mathrm{Cov}[i_{j,\ell_1} , i_{j,j-1}]
%-\mathrm{Cov}[i_{j,\ell_2} , i_{j,j-1}]
\\&=
  \Vsf \left(P_{j,\min\{\ell_1, \ell_2\}},P_{j,\max\{\ell_1, \ell_2\}}\right)
+ \Vsf \left(P_{j,j-1}\right)
\\&
- \Vsf \left(P_{j,j-1},P_{j,\ell_1}\right)
- \Vsf \left(P_{j,j-1},P_{j,\ell_2}\right).
%\\&=\text{\red what does this tell us?}
\end{align}
%Note that, although not explicitly stated so as not to render the notation too heavy, the mean $\bm{\mu}_j$ and the covariance $\bm{V}_j$ depend on $(\alpha_1,\ldots\alpha_K)$. As this dependance is important next, we shall use in rest of this remark $\bm{\mu}_j(\alpha_1,\ldots\alpha_K)$ and $\bm{V}_j(\alpha_1,\ldots\alpha_K)$.
\label{eq:mu and Sigma of i_{j,ell}-i_{j,j-1}}
\end{subequations}

\begin{subequations}
Finally, for independent noises (i.e., block diagonal dispersion matrix), 
%and with an analysis that closely follows the one for the two-user case, 
we obtain that the following second order region is achievable with block-length $n$ and global reliability $\varepsilon$
%for fixed $(\alpha_1,\ldots,\alpha_K)$ as in~\eqref{eq:alphaKuser} is 
\begin{align}
&\bigcup_{ \substack{ \sum_{j\in[K]}\alpha_j \leq 1 \\ \prod_{j\in[K]}(1-\epsilon_j)\geq 1-\varepsilon } }
\bigcap_{j\in[K]} \Big\{(R_1,R_2,\ldots,R_K)\in \mathbb{R}^K_+ : 
\\&\begin{bmatrix}
  R_K+R_{K-1} + \ldots + R_j \\
      R_{K-1} + \ldots + R_j \\
 \vdots \\
                R_{j-1}+ R_j \\
                         R_j \\
\end{bmatrix}                         
\in \bm{\mu}_j(\alpha_1,\ldots\alpha_K)
%\begin{bmatrix}
%  I_{j,K} -I_{j,j-1}\\
% I_{j,K-1} -I_{j,j-1} \\
% \vdots \\
%I_{j,j+1} -I_{j,j-1}\\
%I_{j,j} -I_{j,j-1}\\
%\end{bmatrix}                         
\\&+\frac{1}{\sqrt{n}} \Qsf_\text{\rm inv}\big(\epsilon_j;\bm{V}_j(\alpha_1,\ldots\alpha_K)\big) 
\Big\}
   + O_{\ln(n)/n} \bm{1},
\end{align}
\label{eq:KuserSupFinal}
%\textcolor{purple}{Should this be - $Q_\text{inv}$?} DT: NO, see eq 74

\noindent where the constraint $\sum_{j\in[K]}\alpha_j \leq 1$ represents how power is allocated across private messages and $\prod_{j\in[K]}(1-\epsilon_j)\geq 1-\varepsilon$ how reliability is allocated across receivers.
\end{subequations}

%%%{\red 
%%%QUESTION: What can we say about the K user case?
%%%}
%%%{\magenta The reduction in dispersion for power shell SUP construction versus i.i.d Gaussian is present for the $K$ user case as well. The multi-level construction of SUP for $K$ users in fact amplifies the disparity making a power shell code construction scheme more advantageous.}{\red Add Graphic }
%%%
%%%{\magenta For larger values of $K$ the power split among a large set of users will often result in effectively very low capacity channels to some set of the users. As in the 2-user case, this results in situations where rate splitting (i.e. CCP) will outperform SUP alone. The largest SUP rate region will involve grouping receivers of similar capacity as a single superposition stage and splitting that rate among users for some points on the achievable rate region boundary.}

\begin{rem}[On Per-User Error]
\rm
Without the optimization over $\prod_{j\in[K]}(1-\epsilon_j)\geq 1-\varepsilon$, the region  in~\eqref{eq:KuserSupFinal} is achievable with per-user average error probability bounded by $\epsilon_j$ for receiver $j\in[K]$. With per-user error, all $K!$ superposition coding ordering should be considered.
%%%{\red RIGHT?}
%%%{\red QUESTION: should we add plots here with per-user error rates and considering the various regions obtained by permuting the role of the users?}
%%%{\magenta As in the two user case with Per-User Reliability, [ALLERTON] it is not sufficient to only consider a SUP construction ordering based on SNR only. The best ordering will depend on both the channel conditions and the required reliability of each user.}
\hfill$\square$
\end{rem}

\section{Conclusions}
%In this paper we showed that for the AWGN BC, utilizing a power shell code construction for superposition coding achieves a larger achievable rate region than utilizing an i.i.d Gaussian construction.  This results from the decrease in dispersion the power shell code construction elicits compared to i.i.d construction.  
In this paper we provided achievable and converse second order rate regions for the AWGN BC with both global and per-user reliability constraints.
In addition, for the two-user case, rate splitting is shown to enlarge the achievable region for a large set of channel conditions.
Surprisingly, rate splitting is only required to achieve CCP, that is, to have all information bits encoded into a single codeword.
%when no SUP is active, resulting in the SUP region being simply the union of CCP and SUPNoRS.  These results are shown to
Extensions to the $K$-user case were discussed. We note that our construction utilizes codewords on the power shell, which achieves a lower dispersion than utilizing an i.i.d Gaussian codebook. The second order terms in our achievable and converse regions do not match. Tightening the converse bound and enlarging the achievable bound (by considering for example Marton's coding for the finite blocklength) are part of ongoing work.

%Such a construction via Dirty Paper Coding would eliminate the sum rate constraint in all non-rate splitting points of the SUP boundary. If such a construction has the same dispersion or better than power shell SUP the achievable rate region would be larger than what has been shown here.

%\section*{Acknowledgement:} 
%This work was supported in part by NSF Award 1900911. Any opinions, findings, and conclusions or recommendations expressed in this material are those of the Authors and do not necessarily reflect the views of the NSF.

\appendix

\begin{lem}[Han-type bounds]\label{lemma:han-type bounds}
Similarly to \cite{737512} we have
\begin{subequations}  
\begin{align}  
&\Pr
  \left[
  \frac{1}{n}\ln\frac{W_1^n(\bm{Y}_1|\bm{X})}{Q_{Y_1|X_2}^n(\bm{Y}_1|\bm{X}_2)} > \frac{1}{n}\ln(M_1^\prime) + \gamma
  \right]_{P_{\bm{X}_2,\bm{X}}Q_{Y_1|X_2}^n}   
\\&= \int_{\substack{ \\  \\ \\ (\bm{u},\bm{x},\bm{y}) : \frac{1}{n}\ln\frac{W_1^n(\bm{y}|\bm{x})}{Q_{Y_1|X_2}^n(\bm{y}|\bm{u})} > \frac{1}{n}\ln(M_1^\prime) + \gamma} }  
\hspace*{-4cm}
P_{\bm{X}_2,\bm{X}}(\bm{u},\bm{x}) \ Q_{Y_1|X_2}^n(\bm{y}|\bm{u})\  {\rm d} \bm{u} {\rm d} \bm{x} {\rm d} \bm{y}
\\&\leq \int_{\substack{ \\  \\ \\ (\bm{u},\bm{x},\bm{y}) : \frac{W_1^n(\bm{y}|\bm{x})}{M_1^\prime e^{n\gamma}} > Q_{Y_1|X_2}^n(\bm{y}|\bm{u})}  }
\hspace*{-3cm}
P_{\bm{X}_2,\bm{X}}(\bm{u},\bm{x}) \frac{e^{-n \gamma}}{M_1^\prime} \ W_1^n(\bm{y}|\bm{x})\  {\rm d} \bm{u} {\rm d} \bm{x} {\rm d} \bm{y}
\\&\leq \frac{e^{-n \gamma}}{M_1^\prime}. 
\end{align}   
\label{eq:RN Y1 given X2 Han's bound}
\end{subequations}  
Similarly
\begin{subequations}  
\begin{align}  
&\Pr
  \left[
  \frac{1}{n}\ln\frac{W_1^n(\bm{Y}_1|\bm{X})}{Q_{Y_1}^n(\bm{Y}_1)} > \frac{1}{n}\ln(M_1^\prime M_2^\prime) + \gamma
  \right]_{P_{\bm{X}_2,\bm{X}}Q_{Y_1}^n}
%\\&= \int_{\substack{ \\  \\ \\ (\bm{u},\bm{x},\bm{y}) : \frac{1}{n}\ln\frac{W_1^n(\bm{y}|\bm{x})}{Q_{Y_1}^n(\bm{y})} > \frac{1}{n}\ln(M_1^\prime M_2^\prime) + \gamma}}
%\hspace*{-4cm} 
%P_{\bm{X}_2,\bm{X}}(\bm{u},\bm{x}) \ Q_{Y_1}^n(\bm{y})\  {\rm d} \bm{u} {\rm d} \bm{x} {\rm d} \bm{y} 
%\\&\leq \int_{\substack{ \\  \\  \\ (\bm{u},\bm{x},\bm{y}) : \frac{W_1^n(\bm{y}|\bm{x})}{M_1^\prime M_2^\prime e^{n\gamma}} > Q_{Y_1}^n(\bm{y})} } 
%\hspace*{-3cm}
%P_{\bm{X}_2,\bm{X}}(\bm{u},\bm{x}) \frac{e^{-n \gamma}}{M_1^\prime M_2^\prime} \ W_1^n(\bm{y}|\bm{x})\  {\rm d} \bm{u} {\rm d} \bm{x} {\rm d} \bm{y}
%\\&
\! \! \! \! \! \!
\leq  \frac{e^{-n \gamma}}{M_1^\prime M_2^\prime},
\end{align}
\label{eq:RN Y1 Han's bound}
\end{subequations}  
and
\begin{subequations}  
\begin{align}  
&\Pr
  \left[
  \frac{1}{n}\ln\frac{Q_{Y_2|X_2}^n(\bm{Y}_2|\bm{X}_2)}{Q_{Y_2}^n(\bm{Y}_1)} > \frac{1}{n}\ln(M_2^\prime) + \gamma
  \right]_{P_{\bm{X}_2,\bm{X}}Q_{Y_2}^n}   
%\\&= \int_{\substack{ \\  \\ \\ (\bm{u},\bm{x},\bm{y}) : \frac{1}{n}\ln\frac{Q_{Y_2|X_2}^n(\bm{y}|\bm{u})}{Q_{Y_2}^n(\bm{y})} > \frac{1}{n}\ln(M_2^\prime) + \gamma} }
%\hspace*{-4cm}
%P_{\bm{X}_2,\bm{X}}(\bm{u},\bm{x}) \ Q_{Y_2}^n(\bm{y}) \ {\rm d} \bm{u} {\rm d} \bm{x}  {\rm d} \bm{y}
%\\&\leq \int_{\substack{ \\  \\ \\  (\bm{u},\bm{x},\bm{y}) : \frac{Q_{Y_2|X_2}^n(\bm{y}|\bm{u})}{M_2^\prime e^{n\gamma}} > Q_{Y_2}^n(\bm{y})} }
%\hspace*{-3cm}
%P_{\bm{X}_2,\bm{X}}(\bm{u},\bm{x}) \frac{e^{-n \gamma}}{M_2^\prime} \ Q_{Y_2|X_2}^n(\bm{y}|\bm{u}) \ {\rm d} \bm{u} {\rm d} \bm{x}  {\rm d} \bm{y}
%\\&
\! \! \! \! \! \!
\leq \frac{e^{-n \gamma}}{M_2^\prime}. 
\end{align}
\label{eq:RN Y2 Han's bound}
\end{subequations}  
\end{lem}

\begin{lem}[Constant $K_{0}$]\label{lemma:K0}
\begin{align}
\sup_{\bm{u} \in \mathbb{R}^n, \bm{y} \in \mathbb{R}^n} \frac
   {P_{\bm{Y}_1|\bm{X}_2}(\bm{y}|\bm{u})} 
   {Q_{Y_1|X_2}^n(\bm{y}|\bm{u})} 
   \leq  27~\sqrt{\frac{\pi e}{8}} \frac{1+2 \gamma_1}{\sqrt{1+4 \gamma_1}} =: K_{0}.
\label{eq:RN Y1 given X2}
\end{align}
\end{lem}

{\bf Proof of Lemma~\ref{lemma:K0}.}
Our proof is similar, and leverages the results of \cite{Laneman-MAC}.
Codewords are chosen from the set $\mathcal{D}_n(\rho,P_1,P_2)$.
By the spherical symmetry of the system and by a rotation of the coordinate axis, we can take WLOG %without loss of generality 
\begin{align}
 \bm{x}_2(m_2^\prime)=(0^{n-1},\sqrt{nP_2}).
\end{align}
By the code construction defining $\mathcal{D}_n(\rho,P_1,P_2)$, we have
\begin{align}
 \bm{x}_1(m_1^\prime,m_2^\prime)=(a^{n-1}(m_1^{\prime}),\rho\sqrt{nP_1}),
\end{align}
with $a^{n-1}(m_1^\prime)$ drawn uniformly at random from the power sphere $\mathcal{S}_{n-2}(\sqrt{n(1-\rho^2)P_1})$.
The transmitted codeword is
\begin{align}
%\bm{x}(m_1^\prime,m_2^\prime) =
\bm{x}_1(m_1^\prime,m_2^\prime)+\bm{x}_2(m_2^\prime)=(a^{n-1}(m_1^\prime),\ \xi\sqrt{nP_2}), 
\label{eq:app TX x}
\end{align}
%{\blue PS: I changed $m_1$ to $m_1^\prime$ in right hand sides of two above align environments.}
or equivalently,
\begin{align}
\bm{x}(m_1^\prime,m_2^\prime)-\xi\bm{x}_2(m_2^\prime)=(a^{n-1}(m_1^\prime),0).
\end{align}
Recall $(1-\rho^2) P_1= \alpha P$ from~\eqref{eq:powalpha}.
Therefore, from~\eqref{eq:app TX x}, we see we can decompose $P_{\bm{Y}_j|\bm{X}_2}(\cdot|(0^{n-1},\sqrt{nP_2}))$ (obtained by averaging over the distribution of $a^{n-1}(m_1^\prime)$) into the product of two distributions: 
(i) the first $n-1$ coordinates have the distribution induced by the uniform distribution on the power-sphere at the output of a point-to-point Gaussian channel with average SNR per channel use $\frac{n(1-\rho^2)P_1}{(n-1)\sigma_j^2} = \frac{n}{n-1}\alpha \gamma_j$; and
(ii) the last coordinate is $\Ncal(\xi\sqrt{nP_2}, \sigma_j^2)$.
%Gaussian with mean $\xi\sqrt{nP_2}$ and variance $\sigma_j^2$.
From~\eqref{eq:limitinggaussian Y given U}, the reference distribution $Q_{Y_1|X_2}^n(\cdot|(0^{n-1},\sqrt{nP_2}))$ is jointly Gaussian with mean $\xi \bm{x}_2(m_2^\prime)$ and covariance matrix $(\sigma_j^2+\alpha P)\bm{I}_n$, that is, (i) the first $n-1$ coordinates are i.i.d. $\Ncal(0,\sigma_j^2+\alpha P)$, and (ii) the last coordinate is $\Ncal(\xi\sqrt{nP_2}, \sigma_j^2+\alpha P)$.
Therefore, 
\begin{subequations}
%\begin{align}
%  P_{\bm{Y}_j|\bm{X}_2}&(\bm{y}|(0^{n-1},\sqrt{nP_2}))
%  &= \int \frac{\delta(x_{n} - \sqrt{nP_2} - \rho\sqrt{nP_1},  \Vert x^{n-1} \Vert^2 - n (1-\rho^2) P_1)}{S_{n-1}(\sqrt{n (1-\rho^2) P_1)}} 
%  \frac{1}{(2\pi\sigma_j)^{n/2}} 
%  %\mathrm{e}^{-\frac{(y_n-\sqrt{nP_2}-\rho\sqrt{nP_1})^2 }{2\sigma_j^2}}
%  \mathrm{e}^{-\frac{\Vert y^{n} - x^{n} \Vert^2}{2\sigma_j^2}} \ d x^{n}
%\\&= \frac{1}{\sqrt{2\pi\sigma_j^2}}\mathrm{e}^{-\frac{(y_n-\sqrt{nP_2}-\rho\sqrt{nP_1})^2 }{2\sigma_j^2}} 
%    \times \text{``$P_Y$ of Gaussian point-to-point as in [LANEMAN] with $n \gets n-1, S \gets \frac{n}{n-1}(1-\rho^2)\frac{P_1}{\sigma_j^2}$''}. 
%\\
%  Q^n_{Y_j|X_2}&(\bm{y}|(0^{n-1},\sqrt{nP_2}))
%\\&= \frac{1}{\sqrt{2\pi((1-\rho^2)P_1+\sigma_j^2)}}\mathrm{e}^{-\frac{(y_n-\sqrt{nP_2}-\rho\sqrt{nP_1})^2 }{2((1-\rho^2)P_1+\sigma_j^2)}} 
%     \times \Ncal\left(y^{n-1}; 0^{n-1}, ((1-\rho^2)P_1+\sigma_j^2) \bm{I}_{n-1} \right)
%     %\frac{1}{(2\pi((1-\rho^2)P_1+\sigma_j^2))^{(n-1)/2}} \mathrm{e}^{-\frac{\Vert y^{n-1} \Vert^2}{2((1-\rho^2)P_1+\sigma_j^2)}} 
%\end{align}
%thus 
%{\red DR: do we need something like $n-1\geq 3$?}{\blue PS: I am not sure where you are getting that.  I can see $n\geq3$ to avoid being forced to choose $m^\prime$ from a degenerate distribution.  }
\begin{align}
  &\frac{P_{\bm{Y}_j|\bm{X}_2}(\bm{y}|(0^{n-1},\sqrt{nP_2}))}{ Q^n_{Y_j|X_2}(\bm{y}|(0^{n-1},\sqrt{nP_2}))}
\\&\leq 
  \frac{\Ncal\left(y_n; \xi\sqrt{nP_2}, \sigma_j^2 \right)}
       {\Ncal\left(y_n; \xi\sqrt{nP_2}, \sigma_j^2+\alpha P \right)}
\label{eq:app last coordinate}
\\&\cdot 
  \frac{\Ncal\left(y^{n-1}; 0^{n-1}, (\sigma_j^2 + \frac{n}{n-1}\alpha P) \bm{I}_{n-1} \right)}
       {\Ncal\left(y^{n-1}; 0^{n-1}, (\sigma_j^2 +              \alpha P) \bm{I}_{n-1} \right)}      
\label{eq:app first n-1 coordinate 2}
\\&\cdot \left.
  27~\sqrt{\frac{\pi}{8}} \frac{1+\xi_n}{\sqrt{1+2\xi_n}}
  \right|_{\xi_n := \frac{n}{n-1} (1-\rho^2)\frac{P_1}{\sigma_j^2}}
\label{eq:app first n-1 coordinate 1}
%\\&\leq 1 \cdot \sqrt{e} \cdot  27~\sqrt{\frac{\pi}{8}} \frac{1+2(1-\rho^2)\frac{P_1}{\sigma_j^2}}{\sqrt{1+4(1-\rho^2)\frac{P_1}{\sigma_j^2}}}
\\&\leq 1 \cdot \sqrt{e} \cdot  27~\sqrt{\frac{\pi}{8}} \frac{1+2\gamma_j}{\sqrt{1+4\gamma_j}},
\end{align}
where~\eqref{eq:app last coordinate} is the contribution of the last coordinate, where~\eqref{eq:app first n-1 coordinate 1} is from \cite[Eq. 104]{Laneman-MAC}, and where~\eqref{eq:app first n-1 coordinate 2} is to account for the average SNR per channel use equal to $\frac{n}{n-1}\alpha \gamma_j$ on the first $n-1$ coordinates, as opposed to $\alpha \gamma_j$.
%where in the last line above we used $n=2$.
\label{eq:mess1}
\end{subequations}

\begin{lem}[Constants $K_{j}$'s]\label{lemma:Kj}
\begin{align}
\sup_{\bm{y} \in \mathbb{R}^n} \frac{P_{\bm{Y}_j}(\bm{y})}{Q^n_{Y_j}(\bm{y})}
  &\leq 27\sqrt{\frac{\pi}{8}} \frac{1+\gamma_j}{\sqrt{1+2 \gamma_j}} =: K_{j},  j\in[2].
\label{eq:RN Y1}
\end{align}
\end{lem}

{\bf Proof of Lemma~\ref{lemma:Kj}.}
Let $\gamma_j=P/\sigma_j^2$.
In \cite[Eq. 43]{Laneman-MAC} it was proved that that~\eqref{eq:RN Y1} holds for (a) $P_{\bm{Y}_j}$ is the distribution induced by the uniform distribution on $\mathcal{S}_{n-1}(\sqrt{nP})$ at the output of a point-to-point Gaussian channel with average noise power $\sigma_j^2$, and (b) $Q^n_{Y_j}(\bm{y})$ is the i.i.d. Gaussian distribution with zero mean and variance $\sigma_j^2+P=\sigma_j^2(1+\gamma_j)$.
%In Lemma~\ref{lemma:uni on power sphere} 
In \cite{MACDMS} it was shown that our superposition code construction induces the uniform distribution on $\mathcal{S}_{n-1}(\sqrt{nP})$, and thus~\eqref{eq:RN Y1} holds to our AWGN BC scenario as well.

\begin{lem}\label{lemma:BE}
The multivariate Berry-Essen~\cite[Theorem 11, for $d=3$]{MACDMS}  states that for all convex, Borel measurable subsets of $\mathbb{R}_d$, we have that the constant $B$ in~\eqref{eq:alomostdone} satisfies
\begin{align}
  B &\leq\frac{k_3 \ z}{  \sqrt{n} \ \big(\lambda_\text{min}\big(\bm{V}(\alpha) + \bm{V}(\bm{x}_2,\bm{x})\big)\big)^{3/2}};
\\ k_d &=42d^{1/4}+16 \ \text{from \cite{10.3150/18-BEJ1072} for $d=3$},
\end{align}

where $z :=\frac{1}{n} \sum_{t\in[n]} \mathbb{E} \left[ (\theta_t^T \theta_t)^{3/2} \right]$ for 
\begin{align}
\theta_t &:=
\begin{bmatrix}
 \frac{(1 - N_{2,t}^2) \gamma_2 + 2 \frac{x_t}{\sigma_2} N_{2,t}}{2(1+ \gamma_2)} 
-\frac{(1 - N_{2,t}^2) \alpha \gamma_2 + 2 \frac{x_t-\xi x_{2,t}}{\sigma_2} N_{2,t}}{2(1+\alpha \gamma_2)}
\\
\frac{(1 - N_{1,t}^2) \alpha \gamma_1 + 2 \frac{x_t-\xi x_{2,t}}{\sigma_1} N_{1,t}}{2(1+\alpha \gamma_1)}
\\
\frac{(1 - N_{1,t}^2) \gamma_1 + 2 \frac{x_t}{\sigma_1} N_{1,t}}{2(1+ \gamma_1)} 
\end{bmatrix};
\end{align}
Note, that while~\cite{10.3150/18-BEJ1072} only directly applies to $\mathcal{N}(0,\mathbf{I})$, methods similar to~\cite[Corrollary 8]{Tan+Kosut}  can be applied for general covariance matrix $\bm{V}$.  
\end{lem}

{\bf Proof of Lemma~\ref{lemma:BE}.}
For terms with $N_1$ in $\theta_t^T \theta_t$:
\begin{align*}
  &(a_1 N^2 + b_1 N + c_1)^2 + (a_2 N^2 + b_2 N + c_2)^2
%%%\\&=
%%%   N^4 (a_1^2+a_2^2)
%%%\\&
%%%+  N^3 2(a_1b_1+a_2b_2)
%%%\\&
%%%+  N^2 (b_1^2+2a_1c_1+b_2^2+2a_2c_2)
%%%\\&
%%%+  N \ 2(b_1c_1+b_2c_2)
%%%\\&
%%%+      (c_1^2+c_2^2)
%%%\\&\leq
%%%   N^4 (a_1^2+a_2^2)
%%%\\&
%%%+2|N|^3|a_1b_1+a_2b_2|
%%%\\&
%%%+  N^2 (b_1^2+b_2^2+2|a_1c_1+a_2c_2|)
%%%\\&
%%%+2|N| \ |b_1c_1+b_2c_2|
%%%\\&
%%%+   (c_1^2+c_2^2)
%%%\\&\leq
%%%   N^4 (a_1^2+a_2^2)
%%%\\&
%%%+2|N|^3 \sqrt{a_1^2+a_2^2} \sqrt{b_1^2+b_2^2}
%%%\\&
%%%+  N^2 (b_1^2+b_2^2+2\sqrt{a_1^2+a_2^2}\sqrt{c_1^2+c_2^2})
%%%\\&
%%%+2|N| \sqrt{b_1^2+b_2^2}\sqrt{c_1^2+c_2^2}
%%%\\&
%%%+   (c_1^2+c_2^2)
%%%\\&=
%%%( \sqrt{a_1^2+a_2^2} N^2 + \sqrt{b_1^2+b_2^2} |N| + \sqrt{c_1^2+c_2^2})^2
\\&\leq 
  \max\left(a_1^2+a_2^2, \frac{b_1^2+b_2^2}{4}, c_1^2+c_2^2\right) (|N|+1)^4
\\&
=: f^2 \ (|N|+1)^4.
\end{align*}
For terms with $N_2$ in $\theta_t^T \theta_t$:
\begin{align*}
  &(a_3 N^2 + b_3 N + c_3)^2
%%%\\&\leq(|a_3| N^2 + |b_3| \ |N| + |c_3|)^2
\\&\leq 
  \max(|a_3|^2, (|b_3|/2)^2, |c_3|^2) (|N|+1)^4
%\\&
=: g^2 \ (|N|+1)^4.
\end{align*}
Therefore
\begin{align*}
z &:= \frac{1}{n} \sum_{t\in[n]} \mathbb{E} \left[ (\theta_t^T \theta_t)^{3/2} \right], 
\\&
 \leq \frac{1}{n} \sum_{t\in[n]} \mathbb{E} \left[ \left( f_t^2 (|N_{1,t}|+1)^4 + g_t^2 (|N_{2,t}|+1)^4 \right)^{3/2} \right],
\\&
 \leq \frac{\sqrt{2}}{n} \sum_{t\in[n]} \mathbb{E} \left[ |f_t|^3 (|N_{1,t}|+1)^6 + |g_t|^3 (|N_{2,t}|+1)^6  \right], 
\\&
 = \frac{( 76 + 94 \sqrt{2/\pi})\sqrt{2}}{n} \sum_{t\in[n]} ( |f_t|^3 + |g_t|^3  ),
\\&
 \leq \frac{( 76 + 94 \sqrt{2/\pi})\sqrt{2}}{n} \sum_{t\in[n]} ( |f_t|^2 + |g_t|^2  )^{3/2},
 \end{align*}
since in general, for $0 < r < p$ we have $(\sum_i |x_i|^p)^{1/p} = \Vert\bm{x}\Vert_{p}\leq \Vert\bm{x}\Vert_{r}\leq d^{{\frac {1}{r}}-{\frac {1}{p}}}\Vert\bm{x}\Vert_{p}$, thus for $d=r=2$ and $p=3$
\begin{align}
\sqrt{x^2 + y^2}
&\leq 2^{{\frac {1}{2}}-{\frac {1}{3}}}  (|x|^3 + |y|^3)^{1/3}.
%\\
%\\
%\sqrt{(a_2-b_2)^2 + (a_1)^2 + (b_1)^2}
%  &\leq \sqrt{2|a_2|^2 + 2|b_2|^2 + |a_1|^2 + |b_1|^2}
%\\&\leq \sqrt{2}\sqrt{|a_2|^2 + |b_2|^2 + |a_1|^2 + |b_1|^2}
%\\&\leq \sqrt{2} \ 4^{{\frac {1}{2}}-{\frac {1}{3}}} 
%(|a_2|^3 + |b_2|^3 + |a_1|^3 + |b_1|^3)^{1/3}
\end{align}

\bibliographystyle{IEEEtran}
\bibliography{refs_master,refs_daniela,refs_besma}

\end{document}